%% file: main.tex
\pdfoutput=1
\documentclass[12pt,a4paper]{article}
\usepackage{ifthen} % for conditional statements
\newboolean{pdflatex}
\setboolean{pdflatex}{true} % False for eps figures 

\newboolean{articletitles}
\setboolean{articletitles}{true} % False removes titles in references

\newboolean{uprightparticles}
\setboolean{uprightparticles}{false} %True for upright particle symbols

\def\paperauthors{LHCb collaboration} % Leave as is for PAPER, CONF and FIGURE
\def\paperasciititle{Measurement of the D* longitudinal polarization in B0->D*-tau+nu decays} % Set ASCII title here !! MAKE sure it's only ASCII characters !! 
\def\papertitle{Measurement of the \Dstar longitudinal polarization in \BdToDstTauNu decays} % Latex formatted title
\def\paperkeywords{{High Energy Physics}, {LHCb}} 
\def\papercopyright{\the\year\ CERN for the benefit of the LHCb collaboration}
\def\paperlicence{CC BY 4.0 licence}
\def\paperlicenceurl{https://creativecommons.org/licenses/by/4.0/}

\input{preamble}
\usepackage{longtable} 
\usepackage[]{siunitx}

\makeatletter
\g@addto@macro\bfseries{\boldmath}
\makeatother

\begin{document}

%%%%%%%%%%%%%%%%%%%%%%%%%
%%%%% Title     %%%%%%%%%
%%%%%%%%%%%%%%%%%%%%%%%%%
\renewcommand{\thefootnote}{\fnsymbol{footnote}}
\setcounter{footnote}{1}

% %%%%%%% CHOOSE TITLE PAGE--------
%\onecolumn
%\input{title-LHCb-INT}
%\input{title-LHCb-ANA}
%\input{title-LHCb-CONF}
%\input{title-LHCb-FIGURE}
\input{title-LHCb-PAPER}

%\twocolumn
% %%%%%%%%%%%%% ---------

\renewcommand{\thefootnote}{\arabic{footnote}}
\setcounter{footnote}{0}

%%%%%%%%%%%%%%%%%%%%%%%%%%%%%%%%
%%%%%  Table of Content   %%%%%%
%%%%%%%%%%%%%%%%%%%%%%%%%%%%%%%%
%%%% Uncomment if desired
%\tableofcontents
\cleardoublepage

%%%%%%%%%%%%%%%%%%%%%%%%%
%%%%% Main text %%%%%%%%%
%%%%%%%%%%%%%%%%%%%%%%%%%

\pagestyle{plain} % restore page numbers for the main text
\setcounter{page}{1}
\pagenumbering{arabic}

%% Uncomment during review phase. 
%% Comment before a final submission.
%\linenumbers

%% This is the main body
%% It is useful to have a single file so comemnts are not missed in overleaf.
\input{body}

% Do not include this in any draft (just for information in the template)
%\input{acknowledgements_intro}
% Comment this in for paper drafts; do not include this in analysis note, conference and figure reports
\input{acknowledgements}

%\input{supplementary}

%\input{appendix}

% This should be taken out in the final paper
%\input{supplementary-app}

\addcontentsline{toc}{section}{References}
%\setboolean{inbibliography}{true}
\bibliographystyle{LHCb}
\bibliography{main,standard,LHCb-PAPER,LHCb-CONF,LHCb-DP,LHCb-TDR}

\newpage
\input{Authorship_LHCb-PAPER-2023-020}

\iffalse
The author list for journal publications is generated from the
Membership Database shortly after 'approval to go to paper' has been
given.  It is available at \url{https://lbfence.cern.ch/membership/authorship}
and will be sent to you by email shortly after a paper number
has been assigned.  
The author list should be included in the draft used for 
first and second circulation, to allow new members of the collaboration to verify
that they have been included correctly. Occasionally a misspelled
name is corrected, or associated institutions become full members.
Therefore an updated author list will be sent to you after the final
EB review of the paper.  In case line numbering doesn't work well
after including the authorlist, try moving the \verb!\bigskip! after
the last author to a separate line.

The authorship for Conference Reports should be ``The LHCb
collaboration'', with a footnote giving the name(s) of the contact
author(s), but without the full list of collaboration names.

The authorship for Figure Reports should be ``The LHCb
collaboration'', with no contact author and without the full list 
of collaboration names.
\fi

\end{document}

%% file: preamble.tex
\usepackage[top=1in, bottom=1.25in, left=1in, right=1in]{geometry}

\usepackage{microtype}
\usepackage{lineno}  % for line numbering during review
\usepackage{xspace} % To avoid problems with missing or double spaces after
\usepackage{caption} %these three command get the figure and table captions automatically small

\usepackage{graphicx}  % to include figures (can also use other packages)
\usepackage{color}
\usepackage{colortbl}
\graphicspath{{./figs/}} % Make Latex search fig subdir for figures
\usepackage{amsmath} % Adds a large collection of math symbols
\usepackage{amssymb}
\usepackage{amsfonts}
\usepackage{upgreek} % Adds in support for greek letters in roman typeset

\newcommand*\patchAmsMathEnvironmentForLineno[1]{%
\expandafter\let\csname old#1\expandafter\endcsname\csname #1\endcsname
\expandafter\let\csname oldend#1\expandafter\endcsname\csname
end#1\endcsname
 \renewenvironment{#1}%
   {\linenomath\csname old#1\endcsname}%
   {\csname oldend#1\endcsname\endlinenomath}%
}
\newcommand*\patchBothAmsMathEnvironmentsForLineno[1]{%
  \patchAmsMathEnvironmentForLineno{#1}%
  \patchAmsMathEnvironmentForLineno{#1*}%
}
\AtBeginDocument{%
\patchBothAmsMathEnvironmentsForLineno{equation}%
\patchBothAmsMathEnvironmentsForLineno{align}%
\patchBothAmsMathEnvironmentsForLineno{flalign}%
\patchBothAmsMathEnvironmentsForLineno{alignat}%
\patchBothAmsMathEnvironmentsForLineno{gather}%
\patchBothAmsMathEnvironmentsForLineno{multline}%
\patchBothAmsMathEnvironmentsForLineno{eqnarray}%
}

\usepackage{hyperxmp}

\usepackage[pdftex,
            pdfauthor={\paperauthors},
            pdftitle={\paperasciititle},
            pdfkeywords={\paperkeywords},
            pdfcopyright={Copyright (C) \papercopyright},
            pdflicenseurl={\paperlicenceurl}]{hyperref}
\usepackage[colorinlistoftodos,textsize=scriptsize]{todonotes}

\usepackage[bottom,flushmargin,hang,multiple]{footmisc}

\usepackage[all]{hypcap}  % Internal hyperlinks to floats.

\input{lhcb-symbols-def} % Add in the predefined LHCb symbols
\input{custom-symbols-def}

\usepackage{cite} % Allows for ranges in citations
\usepackage{mciteplus}
\usepackage{cleveref}

%% file: lhcb-symbols-def.tex
\usepackage{xspace} 
\usepackage{upgreek}

\def\lhcb   {\mbox{LHCb}\xspace}

\def\babar  {\mbox{BaBar}\xspace}
\def\belle  {\mbox{Belle}\xspace}

\def\MagUp {\mbox{\em Mag\kern -0.05em Up}\xspace}

\ifthenelse{\boolean{uprightparticles}}%
{

 \def\Peta        {\ensuremath{\upeta}\xspace}

 \def\Ppi         {\ensuremath{\uppi}\xspace}                 
                  
 \def\Prho        {\ensuremath{\uprho}\xspace}                 
                  
 \def\Ptau        {\ensuremath{\uptau}\xspace}

 \def\PDelta      {\ensuremath{\Delta}\xspace}                 
 \def\PXi         {\ensuremath{\Xi}\xspace}                 
 \def\PLambda     {\ensuremath{\Lambda}\xspace}                 
 \def\PSigma      {\ensuremath{\Sigma}\xspace}                 
 \def\POmega      {\ensuremath{\Omega}\xspace}                 
 \def\PUpsilon    {\ensuremath{\Upsilon}\xspace}
 \let\oldPi\Pi
 \def\PPi         {\ensuremath{\oldPi}\xspace}

 \def\PB      {\ensuremath{\mathrm{B}}\xspace}                 
                  
 \def\PD      {\ensuremath{\mathrm{D}}\xspace}

 \def\PK      {\ensuremath{\mathrm{K}}\xspace}

 \def\Pb      {\ensuremath{\mathrm{b}}\xspace}                 
 \def\Pc      {\ensuremath{\mathrm{c}}\xspace}

 \def\Pi      {\ensuremath{\mathrm{i}}\xspace}

 \def\Ps      {\ensuremath{\mathrm{s}}\xspace}

 \def\thebaroffset{0.0em}
}
{

 \def\Peta        {\ensuremath{\eta}\xspace}

 \def\Ppi         {\ensuremath{\pi}\xspace}                 
                  
 \def\Prho        {\ensuremath{\rho}\xspace}                 
                  
 \def\Ptau        {\ensuremath{\tau}\xspace}

 \mathchardef\PDelta="7101
 \mathchardef\PXi="7104
 \mathchardef\PLambda="7103
 \mathchardef\PSigma="7106
 \mathchardef\POmega="710A
 \mathchardef\PUpsilon="7107
 \mathchardef\PPi="7105
                  
 \def\PB      {\ensuremath{B}\xspace}                 
                  
 \def\PD      {\ensuremath{D}\xspace}

 \def\PK      {\ensuremath{K}\xspace}

 \def\Pb      {\ensuremath{b}\xspace}                 
 \def\Pc      {\ensuremath{c}\xspace}

 \def\Pi      {\ensuremath{i}\xspace}

 \def\Ps      {\ensuremath{s}\xspace}

 \def\thebaroffset{0.18em}
}
\newcommand{\offsetoverline}[2][\thebaroffset]{\kern #1\overline{\kern -#1 #2}}%

\makeatletter
\ifcase \@ptsize \relax% 10pt
  \newcommand{\miniscule}{\@setfontsize\miniscule{4}{5}}% \tiny: 5/6
\or% 11pt
  \newcommand{\miniscule}{\@setfontsize\miniscule{5}{6}}% \tiny: 6/7
\or% 12pt
  \newcommand{\miniscule}{\@setfontsize\miniscule{5}{6}}% \tiny: 6/7
\fi
\makeatother

\DeclareRobustCommand{\optbar}[1]{\shortstack{{\miniscule (\rule[.5ex]{1.25em}{.18mm})}
  \\ [-.7ex] $#1$}}

\def\tauon      {{\ensuremath{\Ptau}}\xspace}
\def\taup       {{\ensuremath{\Ptau^+}}\xspace}

\def\squark    {{\ensuremath{\Ps}}\xspace}

\def\cquark    {{\ensuremath{\Pc}}\xspace}

\def\bquark    {{\ensuremath{\Pb}}\xspace}
\def\bquarkbar {{\ensuremath{\overline \bquark}}\xspace}
\def\bbbar     {{\ensuremath{\bquark\bquarkbar}}\xspace}

\def\pion   {{\ensuremath{\Ppi}}\xspace}
\def\piz    {{\ensuremath{\pion^0}}\xspace}
\def\pip    {{\ensuremath{\pion^+}}\xspace}
\def\pim    {{\ensuremath{\pion^-}}\xspace}

\def\rhomeson {{\ensuremath{\Prho}}\xspace}
\def\rhoz     {{\ensuremath{\rhomeson^0}}\xspace}

\def\kaon    {{\ensuremath{\PK}}\xspace}
\def\KorKbar {\kern \thebaroffset\optbar{\kern -\thebaroffset \PK}{}\xspace}
\def\Kz      {{\ensuremath{\kaon^0}}\xspace}

\def\Kp      {{\ensuremath{\kaon^+}}\xspace}
\def\Km      {{\ensuremath{\kaon^-}}\xspace}

\newcommand{\etapr}{\ensuremath{\Peta^{\prime}}\xspace}

\def\Dbar    {{\ensuremath{\offsetoverline{\PD}}}\xspace}
\def\D       {{\ensuremath{\PD}}\xspace}

\def\DorDbar {\kern \thebaroffset\optbar{\kern -\thebaroffset \PD}\xspace}
\def\Dz      {{\ensuremath{\D^0}}\xspace}
\def\Dzb     {{\ensuremath{\Dbar{}^0}}\xspace}
\def\Dp      {{\ensuremath{\D^+}}\xspace}
\def\Dm      {{\ensuremath{\D^-}}\xspace}

\def\DpDm    {\ensuremath{\Dp {\kern -0.16em \Dm}}\xspace}
\def\Dstar   {{\ensuremath{\D^*}}\xspace}

\def\Dstarm  {{\ensuremath{\D^{*-}}}\xspace}

\def\Ds      {{\ensuremath{\D^+_\squark}}\xspace}

\def\Dss     {{\ensuremath{\D^{*+}_\squark}}\xspace}

\def\B       {{\ensuremath{\PB}}\xspace}

\def\BorBbar {\kern \thebaroffset\optbar{\kern -\thebaroffset \PB}\xspace}

\def\Bd      {{\ensuremath{\B^0}}\xspace}

\def\BdorBdbar {\kern \thebaroffset\optbar{\kern -\thebaroffset \Bd}\xspace}

\def\Bs      {{\ensuremath{\B^0_\squark}}\xspace}

\def\BsorBsbar {\kern \thebaroffset\optbar{\kern -\thebaroffset \Bs}\xspace}

\def\Y#1S{\ensuremath{\PUpsilon{(#1S)}}\xspace}

\def\LorLbar     {\kern \thebaroffset\optbar{\kern -\thebaroffset \PLambda}\xspace}

\newcommand{\decay}[2]{\ensuremath{#1\!\to #2}\xspace} 

\def\to                 {\ensuremath{\rightarrow}\xspace}

\def\qsq       {{\ensuremath{q^2}}\xspace}

\def\AT#1     {\ensuremath{A_{\mathrm{T}}^{#1}}\xspace}           % 2

\def\C#1      {\ensuremath{\mathcal{C}_{#1}}\xspace}                       % 9
\def\Cp#1     {\ensuremath{\mathcal{C}_{#1}^{'}}\xspace}                    % 7
\def\Ceff#1   {\ensuremath{\mathcal{C}_{#1}^{\mathrm{(eff)}}}\xspace}        % 9  
\def\Cpeff#1  {\ensuremath{\mathcal{C}_{#1}^{'\mathrm{(eff)}}}\xspace}       % 7
\def\Ope#1    {\ensuremath{\mathcal{O}_{#1}}\xspace}                       % 2
\def\Opep#1   {\ensuremath{\mathcal{O}_{#1}^{'}}\xspace}                    % 7

\newcommand{\aunit}[1]{\ensuremath{\text{\,#1}}}       
\newcommand{\tev}{\aunit{Te\kern -0.1em V}\xspace}
\newcommand{\gev}{\aunit{Ge\kern -0.1em V}\xspace}
\newcommand{\mev}{\aunit{Me\kern -0.1em V}\xspace}
\newcommand{\kev}{\aunit{ke\kern -0.1em V}\xspace}
\newcommand{\ev}{\aunit{e\kern -0.1em V}\xspace}
 
\newcommand{\mevc}{\ensuremath{\aunit{Me\kern -0.1em V\!/}c}\xspace}
\newcommand{\gevc}{\ensuremath{\aunit{Ge\kern -0.1em V\!/}c}\xspace}
\newcommand{\mevcc}{\ensuremath{\aunit{Me\kern -0.1em V\!/}c^2}\xspace}
\newcommand{\gevcc}{\ensuremath{\aunit{Ge\kern -0.1em V\!/}c^2}\xspace}
\newcommand{\gevgevcccc}{\ensuremath{\gev^2\!/c^4}\xspace} % for q^2

\def\mm   {\aunit{mm}\xspace}

\def\fb   {\ensuremath{\aunit{fb}}\xspace}
\def\invfb   {\ensuremath{\fb^{-1}}\xspace}

\def\ps   {\ensuremath{\aunit{ps}}\xspace}

\newcommand{\stat}{\aunit{(stat)}\xspace}
\newcommand{\syst}{\aunit{(syst)}\xspace}

\def\gsim{{~\raise.15em\hbox{$>$}\kern-.85em
          \lower.35em\hbox{$\sim$}~}\xspace}
\def\lsim{{~\raise.15em\hbox{$<$}\kern-.85em
          \lower.35em\hbox{$\sim$}~}\xspace}

\def\sPlot{\mbox{\em sPlot}\xspace}

\def\sqs   {\ensuremath{\protect\sqrt{s}}\xspace}

\def\pt         {\ensuremath{p_{\mathrm{T}}}\xspace}

\def\ptot       {\ensuremath{p}\xspace}

\def\evtgen     {\mbox{\textsc{EvtGen}}\xspace}

\def\geant      {\mbox{\textsc{Geant4}}\xspace}

\def\photos     {\mbox{\textsc{Photos}}\xspace}

\def\pythia     {\mbox{\textsc{Pythia}}\xspace}

\def\tell1  {TELL1\xspace}
\def\ukl1   {UKL1\xspace}

\newcommand{\lhcborcid}[1]{\href{https://orcid.org/#1}{\hspace*{0.1em}\raisebox{-0.45ex}{\includegraphics[width=1em]{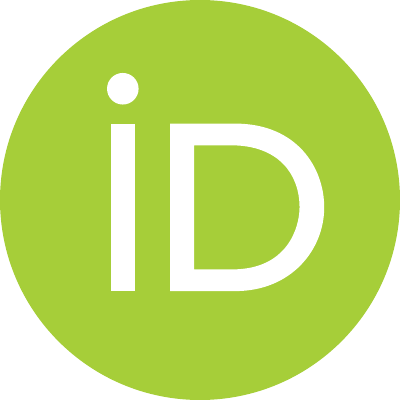}}}}

%% file: custom-symbols-def.tex
\def\pp {\ensuremath{pp}\xspace}

\def\Fldq {\ensuremath{F_L^{\Dstar}(q^2)}\xspace}
\def\aDq {\ensuremath{a_{\theta_D}(q^2)}\xspace}
\def\cDq {\ensuremath{c_{\theta_D}(q^2)}\xspace}
\def\Fld {\ensuremath{F_L^{\Dstar}}\xspace}
\def\aD {\ensuremath{a_{\theta_D}}\xspace}
\def\cD {\ensuremath{c_{\theta_D}}\xspace}

\def\costhetaD {\ensuremath{\cos\theta_{D}}\xspace}

\def\fpol {\ensuremath{f^{\rm pol}}\xspace}
\def\fpollow {\ensuremath{f^{\rm pol}_{\rm{low}\ \it{q^2}}}\xspace}
\def\fpolhigh {\ensuremath{f^{\rm pol}_{\rm{high}\ \it{q^2}}}\xspace}

\def\Nunpol {\ensuremath{N^{\rm unpol}}\xspace}
\def\Nunpollow {\ensuremath{N^{\rm unpol}_{\rm{low}\ \it{q^2}}}\xspace}
\def\Nunpolhigh {\ensuremath{N^{\rm unpol}_{\rm{high}\ \it{q^2}}}\xspace}
\def\Npol {\ensuremath{N^{\rm pol}}\xspace}
\def\Npoltrue {\ensuremath{\Npol_{\rm true}}\xspace}
\def\Nunpoltrue {\ensuremath{\Nunpol_{\rm true}}\xspace}
\def\Ntottrue {\ensuremath{N^{\rm tot}_{\rm true}}\xspace}

\def\PDFunpol {\ensuremath{{\rm PDF}_{\rm unpol}}\xspace}

\def\Dsr {\ensuremath{\D_{\squark}^{(*,**)+}}\xspace}
\def\Dssz {\ensuremath{\D_{\squark0}^{*+}}\xspace}

\def\Dstst {\ensuremath{\D^{**}}\xspace}

\def\etaTopipipi   {\decay{\eta}{\pip\pim\piz}}
\def\etapToetapipi {\decay{\etapr}{\eta\pip\pim}}
\def\etapTorhog    {\decay{\etapr}{\rhoz\gamma}}

\def\BdToDstTauNu  {\decay{\Bd}{\Dstarm\taup\nu_{\tauon}}}
\def\BdToDTauNu  {\decay{\Bd}{\D^{(\ast)-}\taup\nu_{\tauon}}}

\def\BdToDLNu  {\decay{\Bd}{\D^{(\ast)-} \ell^+\nu_{\ell}}}
\def\BdToDststTauNu {\decay{\Bd}{\D^{**-}\taup\nu_{\tauon}}}
\def\BdToDstpipipi      {\decay{\Bd}{\Dstarm\pip\pim\pip}}

\def\BdToDstpipipiX    {\decay{\Bd}{\Dstarm\pip\pim\pip X}}
\def\bbToDstpiX {\decay{\bbbar}{\Dstar\pi X}}

\def\BduToDstDsX {\decay{\B^{0,\pm}}{\Dstar\Ds X}}
\def\BToDstpipipiX {\decay{\B}{\Dstar\pip\pim\pip X}}
\def\tauTopipipi {\decay{\taup}{\pip\pim\pip\nu_{\tauon}}}
\def\tauTopipipiz {\decay{\taup}{\pip\pim\pip\piz\nu_{\tauon}}}
\def\tauTopipipipiz {\decay{\taup}{\pip\pim\pip(\piz)\nu_{\tauon}}}

\def\BdToDstDX {\decay{\Bd}{\Dstar\D^{\pm,0}_{(s)}(X)}}
\def\BdToDstDs {\decay{\Bd}{\Dstarm\Ds}}
\def\BdToDstDss {\decay{\Bd}{\Dstarm\Dss}}
\def\BdToDstDssz {\decay{\Bd}{\Dstarm\Dssz}}

\def\BdToDstDpX {\decay{\Bd}{\Dstarm\Dp(X)}}
\def\BdToDstDspu {\decay{\Bd}{\Dstarm\D^{\prime+}_{\squark1}}}

\def\BdToDstDzX {\decay{\Bd}{\Dstarm\Dz(X)}}

\def\BsToDstDsX {\decay{\Bs}{\Dstarm\Ds(X)}}
\def\BToDstDsX {\decay{\B}{\Dstarm\Ds(X)}}
\def\BToDstDpX {\decay{\B}{\Dstarm\Dp(X)}}
\def\BToDstDzX {\decay{\B}{\Dstarm\Dz(X)}}
\def\BToDstDzK {\decay{\Bd}{\Dstarm\Dz\Kp}}
\def\BToDstDX {\decay{\B}{\Dstarm\D(X)}}

\def\BToDstDpzX {\decay{\B}{\Dstarm(\Dp,\Dz)(X)}}

\def\DsTopipipiX {\decay{\Ds}{\pip\pim\pip(X)}}

\def\DsToetapX  {\decay{\Ds}{\etapr X}}

\def\DzToKpipipi {\decay{\Dz}{\Km\pip\pim\pip}}

\def\DstToDzpi {\decay{\Dstarm}{\Dzb\pim}}
\def\DpToKpipi {\decay{\Dp}{\Km\pip\pip}}
\def\DpToKpipipiz {\decay{\Dp}{\Km\pip\pip(\piz)}}

\def\Cherenkov {Cherenkov\ }

\def\tauola     {\mbox{\textsc{Tauola}}\xspace}
\def\pipi    {\ensuremath{\pip\pim}\xspace}
\def\aone  {\ensuremath{a_1(1260)^+}\xspace}

\def\runo {Run\,1\xspace}
\def\runt {Run\,2\xspace}

%% file: title-LHCb-PAPER.tex
% ===============================================================================
% Purpose: LHCb-PAPER journal paper title page template
% Author: 
% Created on: 2010-09-25
% ===============================================================================

%%%%%%%%%%%%%%%%%%%%%%%%%
%%%%%  TITLE PAGE  %%%%%%
%%%%%%%%%%%%%%%%%%%%%%%%%
\begin{titlepage}
\pagenumbering{roman}

% Header ---------------------------------------------------
\vspace*{-1.5cm}
\centerline{\large EUROPEAN ORGANIZATION FOR NUCLEAR RESEARCH (CERN)}
\vspace*{1.5cm}
\noindent
\begin{tabular*}{\linewidth}{lc@{\extracolsep{\fill}}r@{\extracolsep{0pt}}}
\ifthenelse{\boolean{pdflatex}}% Logo format choice
{\vspace*{-1.5cm}\mbox{\!\!\!\includegraphics[width=.14\textwidth]{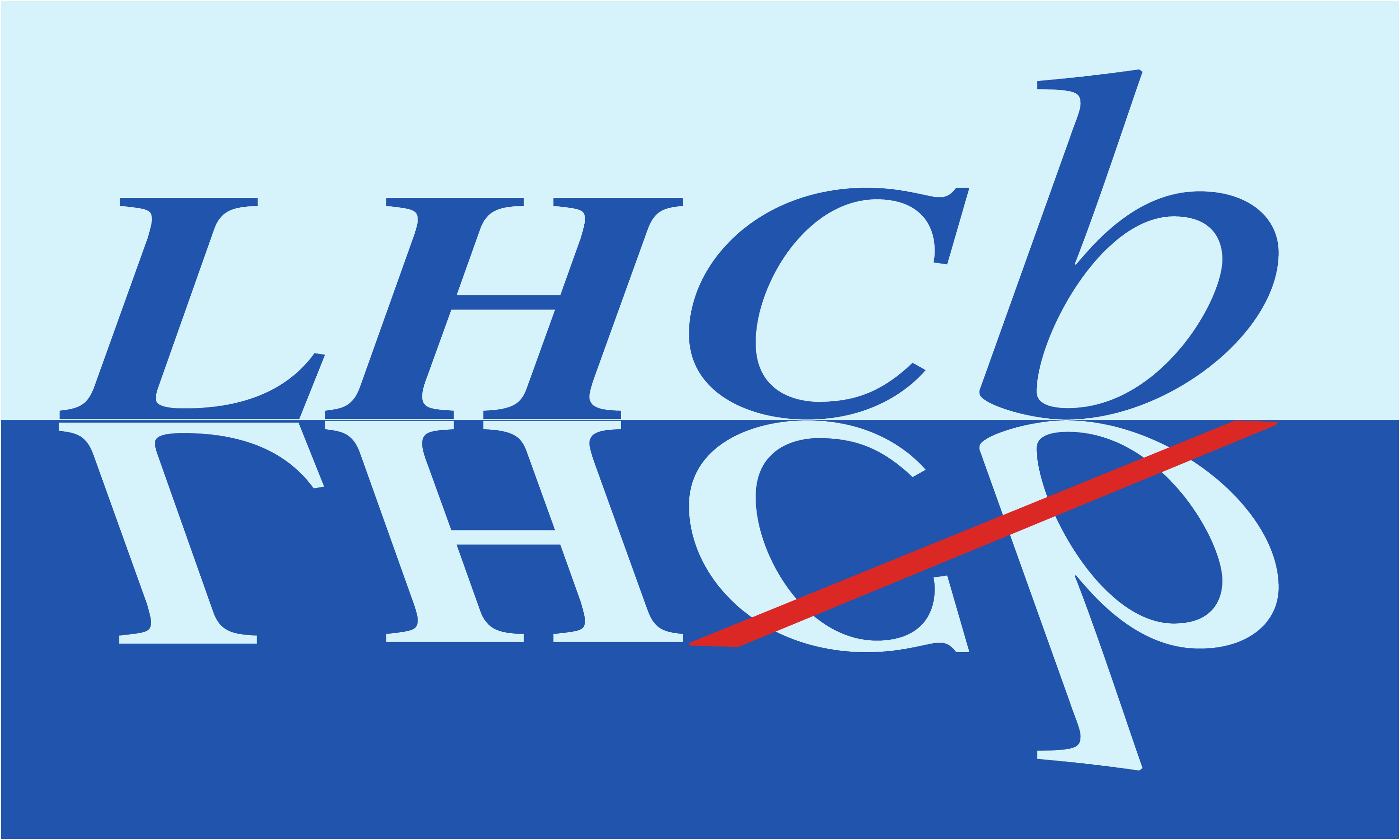}} & &}%
{\vspace*{-1.2cm}\mbox{\!\!\!\includegraphics[width=.12\textwidth]{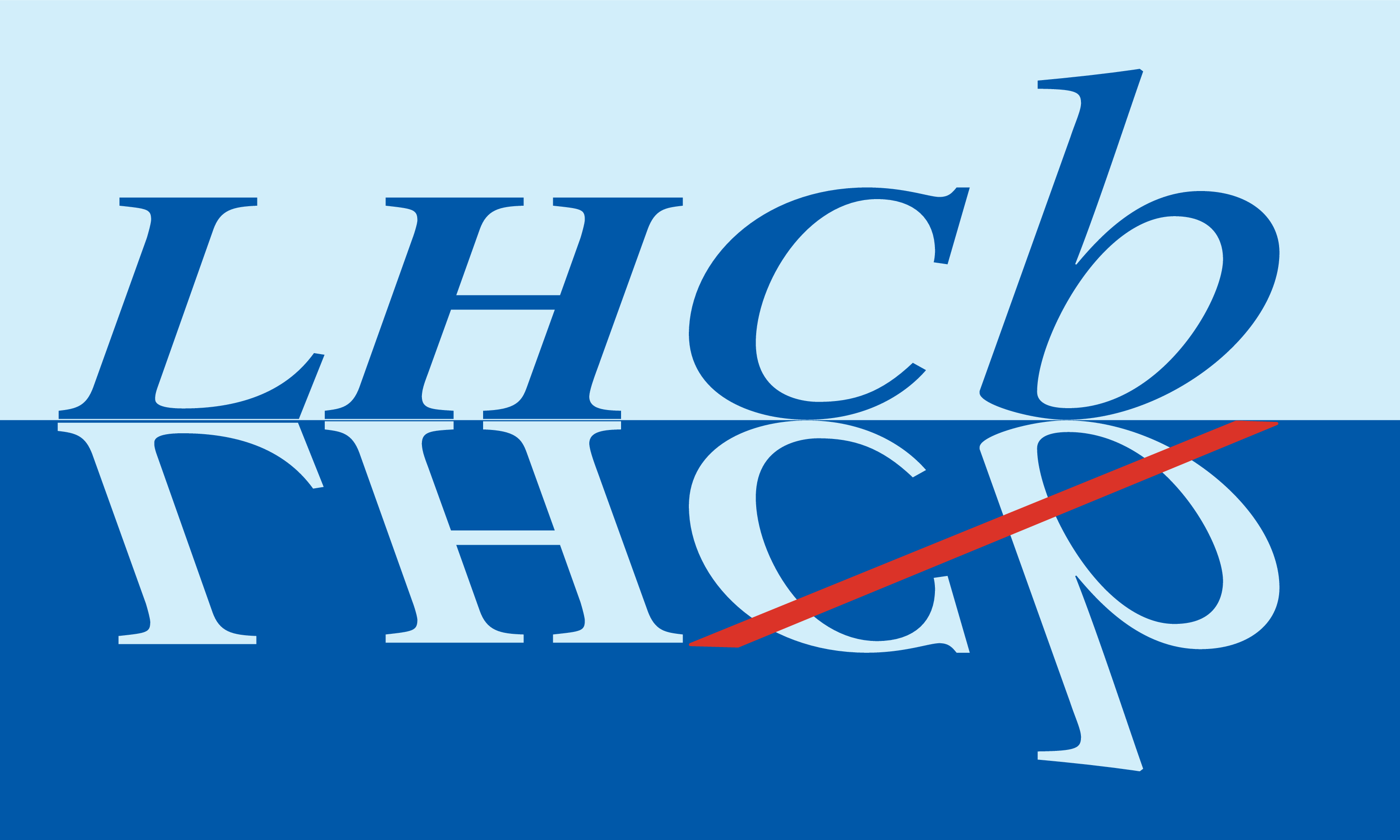}} & &}%
\\
 & & CERN-EP-2023-225 \\  % ID 
 & & LHCb-PAPER-2023-020 \\  % ID 
 %& & \today \\ % Date - Can also hardwire e.g.: 23 March 2010
 & & December 12, 2024 \\
 & & \\
% not in paper \hline
\end{tabular*}

\vspace*{4.0cm}

% Title --------------------------------------------------
{\normalfont\bfseries\boldmath\huge
\begin{center}
% DO NOT EDIT HERE. Instead edit macro in main.tex to keep metadata correct
  \papertitle 
\end{center}
}

\vspace*{1.8cm}

% Authors -------------------------------------------------
\begin{center}
%In the footnote, replace 'paper' by 'Letter' in case of submission to PRL or PLB 
% Edit macro in main.tex to keep metadata correct
\paperauthors\footnote{Authors are listed at the end of this paper.}
\end{center}

\vspace{\fill}

% Abstract -----------------------------------------------
\vspace*{0.5cm}

\begin{abstract}
\vspace*{0.2cm}

 \noindent
 The longitudinal polarization fraction of the \Dstar meson is measured in \BdToDstTauNu decays, where the \tauon lepton decays to three charged pions and a neutrino, using proton-proton collision data collected by the \lhcb experiment at center-of-mass energies of 7, 8 and 13 \tev and corresponding to an integrated luminosity of 5~\invfb. The \Dstar polarization fraction \Fld is measured in two \qsq regions, below and above 7\gevgevcccc, where \qsq is defined as the squared invariant mass of the $\tauon\nu_{\tau}$ system. The \Fld values are measured to be $0.52 \pm 0.07 \pm 0.04 $ and $0.34 \pm 0.08 \pm 0.02$ for the lower and higher \qsq regions, respectively. The first uncertainties are statistical and the second %
 systematic. The average value over the whole \qsq range is: 
 $$\Fld = 0.41 \pm 0.06 \pm 0.03.$$
 These results are compatible with the Standard Model predictions.
\end{abstract}

\vspace*{1cm}

\begin{center}
  Published in Phys. Rev. D 110 (2024) 9, 092007
\end{center}

\vspace{\fill}

{\footnotesize 
% Edit macro in main.tex to keep metadata correct
\centerline{\copyright~\papercopyright. \href{\paperlicenceurl}{\paperlicence}.}}
\vspace*{2mm}

\end{titlepage}

%%%%%%%%%%%%%%%%%%%%%%%%%%%%%%%%
%%%%%  EOD OF TITLE PAGE  %%%%%%
%%%%%%%%%%%%%%%%%%%%%%%%%%%%%%%%

%  empty page follows the title page ----
\newpage
\setcounter{page}{2}
\mbox{~}

%% file: body.tex
\section{Introduction}\label{sec:intro}

Semileptonic decays of \B mesons into final states with a \tauon lepton are a useful test of the Standard Model (SM) and its extensions.
This paper focuses on the \BdToDstTauNu decay,\footnote{The inclusion of charge-conjugate decay modes is implied throughout the paper.} which is theoretically well understood\cite{PhysRevD.92.114022} and is sensitive to new physics (NP) contributions~\cite{PhysRevD.85.094025}.
This decay mode has been studied by both the \babar and \belle experiments~\cite{PhysRevLett.109.101802, PhysRevLett.100.021801, PhysRevD.92.072014, PhysRevLett.99.191807, PhysRevD.94.072007, PhysRevD.97.012004} and the
\lhcb experiment~\cite{LHCb-PAPER-2017-017, LHCb-PAPER-2017-027, LHCb-PAPER-2022-039}.
These measurements aimed to determine the distributions of the most relevant kinematic variables, such as the \tauon polarization, the branching fraction and the ratio \mbox{$\mathcal{R}(\D^{(\ast)}) = \mathcal{B}(\BdToDTauNu)/\mathcal{B}(\BdToDLNu)$, where ${\it \ell}$} denotes an electron or a muon. 
The latest \lhcb analysis of this decay mode obtained a value of $\mathcal{R}(\Dstar)$~=~0.260~$\pm$~0.015~\stat~$\pm$~0.016~\syst~$\pm$~0.012~(ext)~\cite{LHCb-PAPER-2022-052}, performed with proton-proton (\pp) collision data collected at a center-of-mass energy (\sqs) of 13 TeV and corresponding to an integrated luminosity of 2~\invfb.
The current world average for $\mathcal{R}(\Dstar)$ is 0.287 $\pm$ 0.012 and together with the $\mathcal{R}(D)$ measurement shows a tension with the SM at the level of 3.3 standard deviations ($\sigma$)~\cite{HFLAV23}.
Studies of the kinematic and angular distributions, such as the longitudinal \Dstar polarization fraction ($\Fld$), can provide additional sensitivity to possible NP scenarios~\cite{PhysRevD.85.094025}.
For example, the \Fld value could be enhanced by new scalar operators or suppressed by new tensor operators~\cite{PhysRevD.94.094028}.
At present, the only measurement of \Fld was performed by the \belle Collaboration with the result \mbox{\Fld = 0.60 $\pm$ 0.08 \stat $\pm$ 0.04 \syst}~\cite{belle_dstarpol}.
This value is compatible with the SM expectations
~\cite{PhysRevD.92.114022, PhysRevD.87.034028, PhysRevD.95.115038}.
The most recent theoretical predictions, computed with different model assumptions and QCD calculations, converge to a value in a range between 0.43 and 0.46~\cite{PhysRevD.98.095018,Colangelo:2018cnj,Bhattacharya_2019,Bordone_2020,ray2023test,martinelli2023updates}.

This paper presents the first measurement of the longitudinal \Dstar polarization fraction by the \lhcb experiment. It uses a sample of \BdToDstTauNu decays
where the \Dstarm meson decays into the \Dzb\pim final state and the \taup lepton decays to three charged pions and a tauonic neutrino, with the potential presence of an additional neutral pion, \textit{i.e.} \tauTopipipipiz.
The measurement is performed using \pp collision data collected by the \lhcb experiment during 2011--2012 at \sqs = 7 and 8 TeV (\runo) and during 2015--2016 at \sqs = 13 TeV (\runt),
corresponding to an integrated luminosity of 3~\invfb and 2~\invfb, respectively.
The doubly differential decay rate of \DstToDzpi decays,
\begin{equation}\label{eq:decay_rate}
\frac{1}{\Gamma}\frac{\mathrm{d}^2\Gamma}{\mathrm{d}\qsq\mathrm{d}\cos\theta_D} = \aDq + \cDq\cos^2\theta_D,
\end{equation}
depends on the angular distribution of the decay products through $\theta_D$, the angle between the \Dzb meson direction and the direction opposite to the \Bd meson in the \Dstar rest frame. The coefficients of the angular distribution,
\aDq and \cDq, encapsulate the hadronic effects and the fundamental couplings~\cite{becirevic2019lepton} and depend on \qsq, the squared invariant mass of the $\tauon\nu_{\tau}$ system computed as the squared difference between the \B and the \Dstar momenta. The longitudinal polarization \Fldq depends on the angular coefficients as
\begin{equation}\label{eq:fld}
\Fldq = \frac{\aDq + \cDq}{3\aDq + \cDq}
\end{equation}
and also depends on \qsq.
The \qsq computation is performed correcting the \B momentum for the missing neutrinos, achieved by measuring the \B and \tauon line-of-flight vectors\footnote{The line-of-flight vector is the vector pointing from a particle's production vertex to its decay vertex.}
and using the known \B and \tauon masses. The complete reconstruction method was described in detail in Ref.~\cite{LHCb-PAPER-2017-027}.
In this paper, the integrated values of \aDq and \cDq in the \qsq region below and above 7~\gevgevcccc, named simply as \aD and \cD, are provided. From those, the corresponding integrated \Fld values are determined.
These results are combined to determine the average integrated values of \aD, \cD and \Fld in the whole \qsq region.

The work described in this paper is organized as follows: the \lhcb detector and the simulation are described in Section~\ref{sec:lhcb}, the event selection is discussed in  
Section~\ref{sec:selection}, the description of the signal mode is detailed in Section~\ref{sec:signal}, the study of the dominant background components is reported in  Section~\ref{sec:backgrounds}, the final fit and the estimation of \Fld are discussed in  Section~\ref{sec:fit}, 
the various systematic sources affecting this measurement are listed and described in  Section~\ref{sec:systematics}, and the final results are given in  Section~\ref{sec:results}.

\section{Detector and simulation}\label{sec:lhcb}

The \lhcb detector~\cite{LHCb-DP-2008-001,LHCb-DP-2014-002} is a single-arm forward
spectrometer covering the \mbox{pseudorapidity} range $2<\eta <5$,
designed for the study of particles containing \bquark or \cquark
quarks. The detector includes a high-precision tracking system
consisting of a silicon-strip vertex detector surrounding the $pp$
interaction region~\cite{LHCb-DP-2014-001}, a large-area silicon-strip detector located
upstream of a dipole magnet with a bending power of about
$4{\mathrm{\,Tm}}$, and three stations of silicon-strip detectors and straw
drift tubes~\cite{LHCb-DP-2013-003,LHCb-DP-2017-001}
placed downstream of the magnet.
The tracking system provides a measurement of the momentum, \ptot, of charged particles with
a relative uncertainty that varies from 0.5\% at low momentum to 1.0\% at 200\gevc.
The minimum distance of a track to a primary $pp$ collision vertex (PV), the impact parameter, 
is measured with a resolution of $(15+29/\pt)\ \si{\micro m}$,
where \pt is the component of the momentum transverse to the beam, in\,\gevc.
Different types of charged hadrons are distinguished using information
from two ring-imaging \Cherenkov detectors~\cite{LHCb-DP-2012-003,LHCb-DP-2017-001}. 
Photons, electrons and hadrons are identified by a calorimeter system consisting of
scintillating-pad and preshower detectors, an electromagnetic calorimeter and a hadronic calorimeter. Muons are identified by a system composed of alternating layers of iron and multiwire proportional chambers~\cite{LHCb-DP-2012-002, LHCb-DP-2017-001}.
The online event selection is performed by a trigger~\cite{LHCb-DP-2012-004, LHCb-DP-2017-001}, 
which consists of a hardware stage, based on information from the calorimeter and muon
systems, followed by a software stage, which applies a full event
reconstruction.

At the hardware trigger stage, events are required to have a muon with high \pt or a
hadron, photon or electron with high transverse energy in the calorimeters. For hadrons,
the typical transverse energy threshold is 3.5\gev.
The software trigger requires a two-, three- or four-track  secondary vertex with a significant displacement from any primary $pp$ interaction vertex. 
At least one charged particle  must have a transverse momentum $\pt > 1.6\gevc$ and be inconsistent with originating from a PV.
A multivariate algorithm~\cite{BBDT,LHCb-PROC-2015-018} is used for the identification of secondary vertices consistent with the decay of a \bquark hadron.

Simulation is required to model the effects of the detector acceptance and the imposed selection requirements. In the simulation, $pp$ collisions are generated using
\pythia~\cite{Sjostrand:2007gs,*Sjostrand:2006za} 
with a specific \lhcb configuration~\cite{LHCb-PROC-2010-056}.
Decays of unstable particles are described by \evtgen~\cite{Lange:2001uf}, in which final-state
radiation is generated using \photos~\cite{davidson2015photos}.
The signal events are generated forcing the \tauon lepton to decay in the $3\pi\nu_{\tau}$ and $3\pi\piz\nu_{\tau}$ final states using the 
resonance chiral Lagrangian model~\cite{PhysRevD.88.093012} modeled with the \tauola~\cite{Davidson_2012} package, tuned according to the results from the \babar Collaboration~\cite{Nugent_2014}.
The interaction of the generated particles with the detector, and its response,
are implemented using the \geant toolkit~\cite{Allison:2006ve, *Agostinelli:2002hh} as described in Ref.~\cite{LHCb-PROC-2011-006}. 
The underlying $pp$ interaction is reused multiple times, with an independently generated signal decay for each event~\cite{LHCb-DP-2018-004}.

The simulated samples used in this analysis are assigned sets of weights in order to match relevant distributions in the data as closely as possible. These weights account for the \BdToDstTauNu form factors, the transverse momentum and $\eta$ distributions of the \B meson, the $3\pi$ vertex position uncertainty, and, for certain decay modes, the \Ds decay model.
The weights are determined in the initial phase of the analysis and are included in all the subsequent steps. The effect of the weights is validated through specific control samples, such as the \BdToDstpipipi decay.

\section{Candidate selection}\label{sec:selection}

The reconstruction and selection of the \BdToDstTauNu decays are performed following the same strategy described in Refs.~\cite{LHCb-PAPER-2017-027, LHCb-PAPER-2022-052}.
Each signal candidate is built combining a \Dstar meson, decaying as \DstToDzpi, and a 3$\pi$ system originating from a common vertex.
The 3$\pi$ vertex is required to be detached from the \Bd vertex due to the  lifetime of the \tauon lepton.
This requirement suppresses the large background with the 3$\pi$ system originating directly from the \Bd decay, referred to as "prompt-decay" background hereafter. Decays of the form \BdToDstDX, characterized by a signal-like topology, remain the most important background source after this step.

\subsection{Event reconstruction and initial selection}

An initial selection is applied to suppress the prompt and combinatorial backgrounds.
The $\tauon$ candidates are reconstructed from three tracks, identified as pions, with transverse momentum $\pt > 250$\mevc.
Since the $3\pi$ vertex is expected to be well separated from the corresponding PV, this distance is required to be 10 times greater than its uncertainty. 
The radial distance between the $3\pi$ vertex and the beam center in the transverse plane is required to be within the range [0.2, 5.0]~mm to avoid pion triplets coming from a PV or from secondary interactions.

The \Dz meson, which originates from a \Dstar meson decay, is reconstructed in the $\Km\pip$ mode selecting tracks identified as a kaon and a pion, each with momentum $p>2$\gevc and transverse momentum $\pt > 250$~\mevc. In addition, the kaon-pion system must have an invariant mass in the range [1840, 1890]~\mevcc and have $\pt > 1.2$~\gevc.
The \Dstar meson is obtained combining the \Dz candidate and a track, identified as a pion (slow pion), with \mbox{$\pt > 110$~\mevc} and selecting only the combinations with a difference between the \Dstar and \Dz candidate invariant mass in the range [143, 148]~\mevcc. The combinatorial background is further reduced by requiring that the \Dz and the \tauon candidates originate from the same PV, chosen as the PV that fits best to the flight direction of the \B candidate.

The combinations with a \Dstar-\Dz mass difference or a \Dz mass outside the specified ranges are used to study the combinatorial background, as discussed in Section~\ref{sec:backgrounds}.
Finally the \Bd events are built from the combination of the \Dstar and the \tauon candidates.

To suppress the background in which a kaon is misidentified as a pion or vice versa, all the kaon and pion candidates are required to be correctly identified using information provided by the particle identification (PID) system. The chosen requirements were already studied and applied in the previous \lhcb analyses~\cite{LHCb-PAPER-2017-027, LHCb-PAPER-2022-052} which were proven to be effective in rejecting background sources like \BdToDstDpX, where the \Dp decays as \DpToKpipipiz.

Finally, because of the presence of neutrinos in the final state,  the invariant mass of the $\Dstar3\pi$ system is required to be below 5100~\mevcc while the \tauon candidate is required to have an invariant mass lower than 1600~\mevcc, in order to suppress a fraction of  double-charm background.
A full list of all the requirements described in this section is reported in Table~\ref{tab:signal_sel}.

The resolution of the \costhetaD variable is determined using a simulated sample of \BdToDstTauNu decays and looking at the distribution of $({\costhetaD}^{\rm reco}-{\costhetaD}^{\rm true})/{\costhetaD}^{\rm true}$, where ${\costhetaD}^{\rm reco}$ and ${\costhetaD}^{\rm true}$ stand for the reconstructed and generated \costhetaD value, respectively. 
The resulting distribution is centered on zero with a standard deviation of 0.57.

\begin{table}[tb]
  \centering
  \caption{List of all the requirements included in the signal selection.}
  \begin{tabular}{ll}
    Variable & Requirement \\
    \hline
    $\pt$ of \tauon decay product & $> 250$~\mevc \\
    $[z(3\pi)-z(\mathrm{PV})]$/error & $> 10$ \\
    Radial distance $3\pi$ system & $[0.2, 5.0]$~\mm \\
    $m(3\pi)$ & $<1600$~\mevcc \\
    $p$ of \Dz decay product & $> 2$~\gevc \\
    $\pt$ of \Dz decay product & $> 250$~\mevc \\
    $m(K\pi)$ & [1840,1890]~\mevcc  \\
    \pt of the \Dz meson & $> 1.2$~\gevc \\
    $\pt$ of the slow pion & $>110$~\mevc  \\
    $\Delta m = m(K\pi\pi) - m(K\pi)$ & [143,148]~\mevcc \\
    $m(\Dstar3\pi)$ & $< 5100$~\mevcc \\
    $\mathrm{PV}(K\pi)$ & = $\mathrm{PV}(3\pi)$ \\
  \end{tabular}
  \label{tab:signal_sel}
\end{table}

\subsection{Vertex detachment requirements}

The prompt candidates represent the dominant source of background.
The baseline strategy for removing this background is to select only the candidates where
the distance between the \Bd and the $3\pi$ vertices along the beam direction [$\Delta z \equiv z(3\pi) - z(\Bd)$] is larger than 4 times its uncertainty ($\sigma_{\Delta z}$).
This requirement rejects 99.95\% of the prompt background in the \runo data sample. 
In \runt data the rejection is lower (around 90\%) due to the larger uncertainty on the vertex reconstruction,\footnote{The larger uncertainty on the vertex reconstruction in \runt data is due to radiation damage in the Vertex-Locator during the first LHC running period.} which increased by a relative 10\% with respect to the \runo value.
In order to compensate for this uncertainty degradation, a multivariate algorithm, based on a boosted decision tree (BDT)~\cite{Breiman,AdaBoost}, is trained with the TMVA package integrated in ROOT~\cite{TMVA4} taking as input the positions of the \Dz, \Bd and $3\pi$ vertices, their corresponding uncertainties, the invariant mass of the $3\pi$ system and the angle between the $\tauon$ momentum and the line of flight from the $\tauon$ origin to the decay vertex.
In the training phase the signal is represented by the simulated \BdToDstTauNu decay while the background is represented by a simulated \bbToDstpiX sample.
The final requirements on $\Delta z$ and the BDT response are optimized on \runt data in order to maintain the same prompt background rejection as in the \runo data.

\subsection{Isolation requirements}\label{sec:selection:isolation}

After the vertex detachment requirements, the dominant background source arises from the {\BToDstDX} modes. These events are typically characterized by the presence of additional final-state charged particles, compared to the signal mode. 
The main contributions are from the \BToDstDzK, \BToDstDpX and \BToDstDsX decays where the \Dz meson decays into four charged particles and the \Dp or \Ds meson decays into five charged particles.
The rejection of this background is based on the strategies used in Refs.~\cite{LHCb-PAPER-2017-027, LHCb-PAPER-2022-052}, which exploit a dedicated algorithm to determine the isolation of each track in the signal decay chain from all the other tracks in the event.
This requirement preserves around 80\% of the signal candidates while rejecting between 82\% and 95\% of the various partially reconstructed backgrounds.

\subsection{Combinatorial background suppression}

After applying the requirements described above, the expected level of combinatorial background in the \runo sample is below 1\%. In the \runt data contamination is higher ($\sim 5\%$), since the average number of particles produced at the PVs is larger due to the increased center-of-mass energy.
Therefore, a BDT classifier is trained with the aim to further reduce this background in \runt.
The input variables of the BDT classifier include the transverse momentum, the pseudorapidity, the impact parameter and the vertices of the \Dstar and \tauon candidates.
A requirement on the BDT output rejects 75\% of the remaining combinatorial background while retaining 77\% of the signal.

\subsection{\texorpdfstring{\BToDstDsX}{B->D*DsX} background suppression}\label{sec:selection:antiDs}

The additional suppression of the \BToDstDsX background is based on a BDT classifier trained on simulated samples. The input features are related to the different dynamics of the 3\pion system in \tauon and \Ds decays and the isolation of the final state with respect to neutral particles. An important difference lies in the fact that the \tauTopipipi mode decays via the intermediate \aone and $\rhoz\pip$ states.
On the other hand, the \Ds meson decay in the $3\pi(X)$ system mainly occurs via the \etaTopipipi and \etapToetapipi resonances with just a small contribution from the $X\aone$ mode, where $X$ stands for a $\eta$, \etapr, $K^0$, $\omega$ or $\phi$ resonance. The resulting differences in both the minimum and maximum value of the \pipi invariant mass provide significant discrimination.
Furthermore, the performance of the BDT classifier is improved by including isolation variables and the information of the energy deposited in the calorimeter, considering that the \Ds decays are often characterized by additional charged and neutral particles in the final state.
The training is performed on a sample of simulated \BdToDstTauNu decays as signal proxy and simulated \BToDstDsX decays for the background.
The background rejection is about 40\%  for a signal efficiency around 97\%.
The output of the BDT classifier is used as a variable in the final fit because of its importance in discriminating the \BToDstDsX decays background.

\section{Signal description}\label{sec:signal}

The \Fld value is calculated from the \aD and \cD parameters extracted from a binned maximum-likelihood fit to data. The fit uses four-dimensional templates in terms of the following variables:
\begin{itemize}
    \item \costhetaD: defined in Section~\ref{sec:intro};
    \item \qsq: defined in Section~\ref{sec:intro};
    \item $t_{\tauon}$: decay time of the \tauon lepton, taking into account the corrections due to the missing neutrino;
    \item anti-\Ds BDT output: the response of the BDT classifier trained to suppress the background due to the \mbox{\BToDstDsX} decay, defined in Section~\ref{sec:selection:antiDs}.
\end{itemize}

The value of \Fld is measured in two different \qsq regions, below and above 7~\gevgevcccc, and considering six bins of \costhetaD in each region. The binning scheme for the other two variables differs for the \runo and \runt samples, due to their different luminosities. The $t_{\tauon}$ distribution, covering the range [0, 2]\,\ps, is split into four and six bins for the \runo and \runt data sets, respectively. 
Similarly, either six or eight bins are used for the anti-\Ds BDT distribution.

The \aD and \cD parameters are determined by splitting the signal simulated template into two components: one is completely unpolarized while the other is fully polarized. 
With this separation, the \aD and \cD parameters become directly proportional to the yields of the unpolarized and polarized components, respectively, as discussed in Section~\ref{sec:fit}.

Weights are assigned to the simulated events such that the decay kinematics follow the Caprini-Lellouch-Neubert (CLN) parametrization of the form factors~\cite{Caprini_1998}. 
This procedure is performed by means of the Hammer tool~\cite{hammer} using the dedicated RooHammerModel~\cite{RooHammer} interface developed by the \lhcb Collaboration.

The simulated signal \costhetaD distribution is corrected for the reconstruction and selection efficiency and bin migration due to differences between the generated and reconstructed variables.
The reconstruction efficiency correction is determined as the ratio of the generated \costhetaD distribution, after the full signal selection, to the corresponding distribution at generation level in bins of generated \qsq; the bin migration correction is determined as the ratio between the \costhetaD distribution in bins of \qsq considering the reconstructed and generated quantities, after the full signal selection. The ratio between these \costhetaD distributions is computed with the nonuniform distribution of the events in each bin taken into account.

\begin{figure}[tb]
\centering
\includegraphics[width=0.65\textwidth]{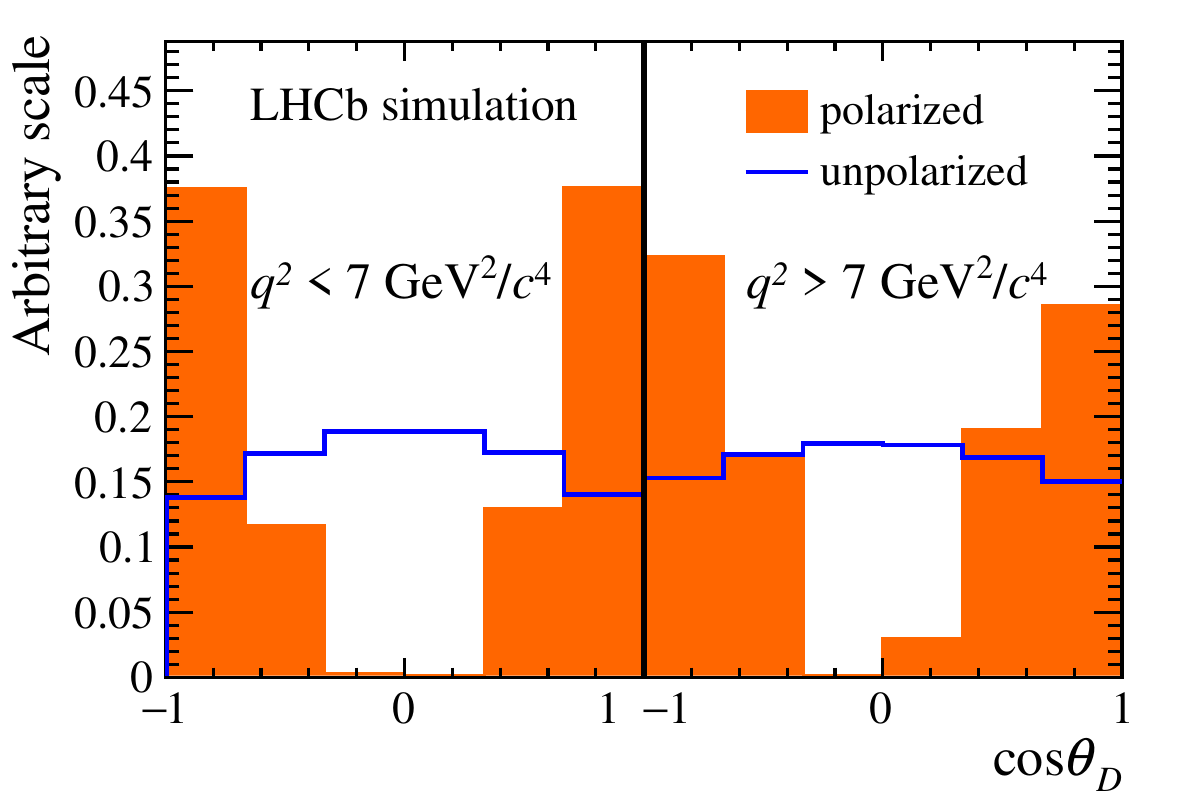}
\caption{Signal distributions of the \costhetaD variable in the two \qsq regions for the unpolarized and polarized components in the \runt simulated sample. Similar distributions are calculated for the \runo sample. The template for the unpolarized component, which is uniform at generation level,  shows clearly the distortion caused by the effect of the reconstruction and the signal selection.}
\label{fig:sigPol}
\end{figure}

The signal \costhetaD distribution in the two \qsq regions for the unpolarized and polarized components, after applying the full signal selection and the form-factor and correction weights, is depicted in Figure~\ref{fig:sigPol} for the \runt sample.
Similar distributions are obtained also from the \runo sample and are used to define the signal probability density functions (PDFs) in the final fit.
The reconstruction effect leads to a variation of the average polarization in both the low- and high-\qsq region by a relative 12\% and 3\%, respectively. Both polarized and unpolarized components are affected by this effect, which is clearly visible in the latter one where the expected flat distribution is distorted.

The total efficiency  for the $\BdToDstTauNu$ mode is estimated from simulation, after corrections are applied. It is evaluated separately for each data-taking period and in each \qsq bin taking into account the effects due to the selection, reconstruction and \qsq bin migration.
These efficiencies are used in the final step of the analysis for correcting the observed signal yields when evaluating the \Fld value.

\section{Background description}\label{sec:backgrounds}

After applying the signal selection,  the residual background consists mainly of events with a charm hadron decay: \BToDstDX decays, where \D can be, in order of importance, a \Ds, \Dp or \Dz meson.
PDFs for these background sources are obtained from simulation.
The corresponding distributions in \runo and \runt data are used to validate and correct the simulation.
Two different control samples are used to validate the simulated \Ds background: the first is based on a pure sample of \DsTopipipiX decays used to correct the branching fractions relevant to \Ds meson production in the generation phase; the second consists of various \BToDstDsX decays and aims to constrain the relative fractions between these components in the final fit.

\subsection{\texorpdfstring{\BToDstDpzX}{B->D*(Dz, Dp)(X)} control sample}

\begin{figure}[tb]
\centering
 \includegraphics[width=0.49\textwidth]{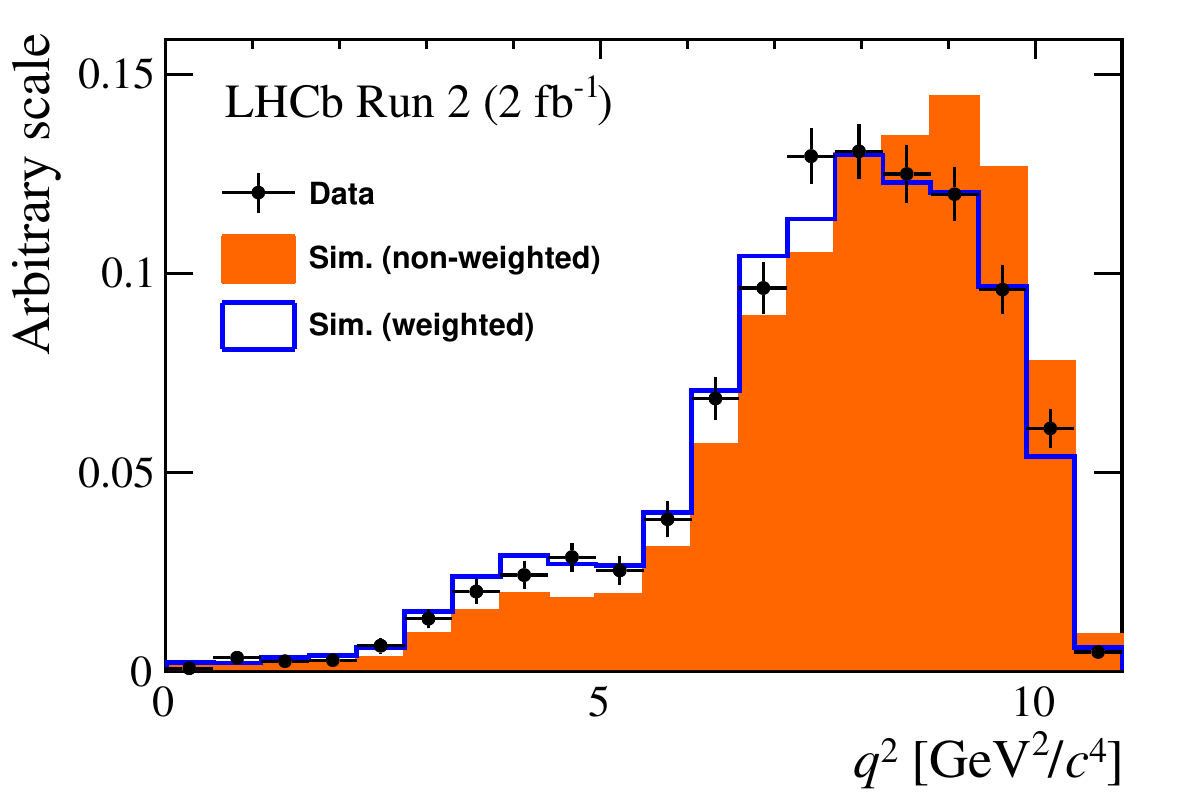} 
 \includegraphics[width=0.49\textwidth]{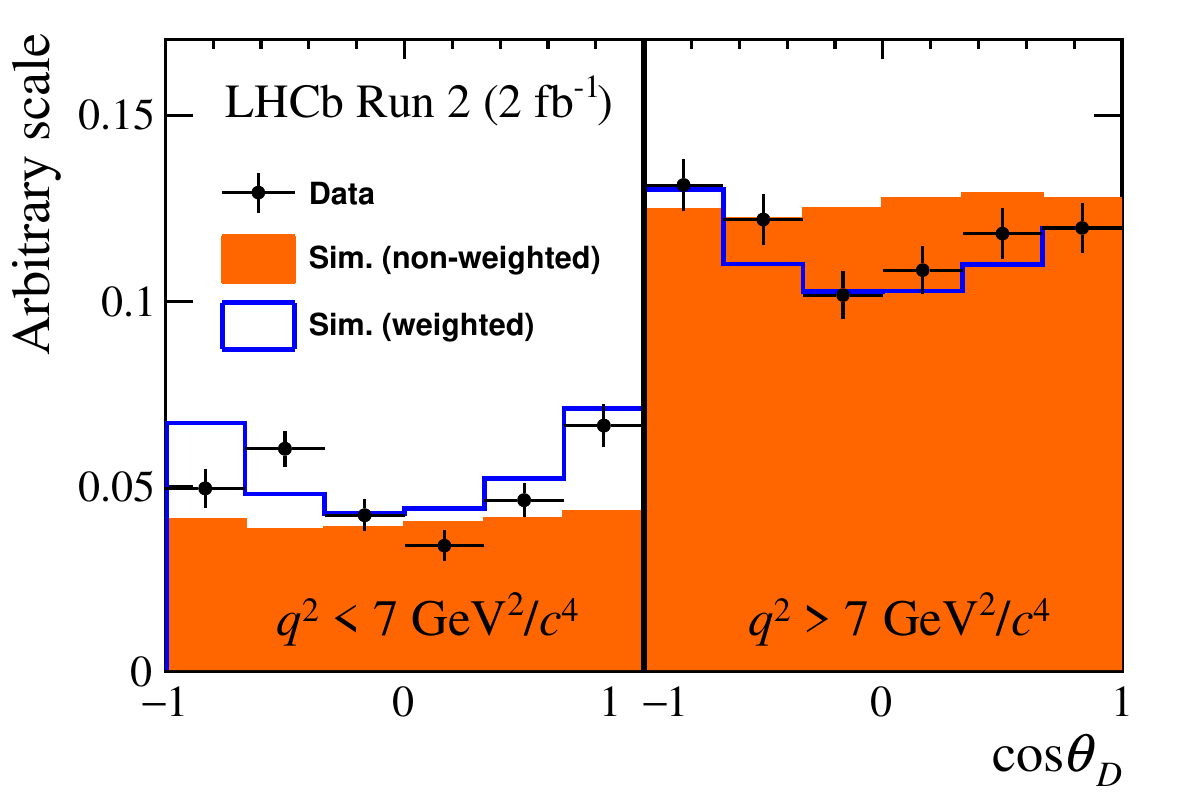} \\
 \caption{Simulated (left) \qsq and (right) \costhetaD distributions for the \BToDstDzX control sample before and after the correction for the data-simulation differences in the \runt weighted data sample. The \BToDstDzX sample is selected using the isolation algorithm to search for an extra kaon around the $3\pi$ vertex.}
 \label{fig:Dcorr}
\end{figure}

The decays \BToDstDpX and \BToDstDzX can contribute to the background when the \Dp meson decays into two charged pions and a kaon, misidentified as a pion, or when a \Dp or  \Dz meson decays into three charged pions and an additional particle.
For the $K\to\pi$ misidentification from \Dp background, a control sample is obtained by inverting the PID requirement on the negatively charged pion and reconstructing the \DpToKpipi decay. 
For the $\Dz\to 3\pi X$ background, a control sample is obtained by reconstructing the \DzToKpipipi decay using the isolation algorithm (described in Section~\ref{sec:selection:isolation}) to search for the extra kaon around the $3\pi$ vertex.
In both cases the \sPlot technique~\cite{Pivk:2004ty} is employed to statistically subtract the combinatorial background and obtain a pure distribution of the decay mode of interest.
The \qsq distribution shows some discrepancies between data and simulation due to the imperfect
modeling of the inclusive $3\pi$ decays. 
A correction is determined as the ratio of the data and simulated \qsq distributions and is applied to the final templates to account for this effect.
Similarly, the \costhetaD distribution is not well modeled in the inclusive simulation and is therefore corrected in order to match the one observed in data. 
The data and simulated \qsq and \costhetaD distributions before and after the corrections for the \BToDstDzX decays are depicted in Figure~\ref{fig:Dcorr}. Similar distributions are obtained for the \BToDstDpX control sample.

\subsection{\texorpdfstring{\DsTopipipiX}{Ds->3pi(X)} decay model}

As mentioned in Section~\ref{sec:selection:antiDs}, the kinematic distributions of the three pions in the final state differ between \tauon and \Ds decays. Special care must be given to \Ds decays that have a \rhoz in the final state, such as the decay through the \etapr intermediate state and the subsequent \etapTorhog decay, with the following branching fractions: $\mathcal{B}(\DsToetapX) =  (10.3 \pm 1.4)\%$ and $\mathcal{B}(\etapTorhog) =  (29.4 \pm 0.4)\%$. 
This case corresponds to kinematic distributions of the three pions that are similar to those of the signal decays.
The aim of the study described in this section is to adjust  the branching fractions of the various resonances generated in the simulated samples to better match the data.
A dedicated control sample enriched in these decays is used, selected using the same criteria as for the signal selection described in Section~\ref{sec:selection} but inverting the requirement on the anti-\Ds BDT output.
The different \Ds modes can be separated in three main categories:
\begin{itemize}
\item \Ds decays where at least one pion comes from an $\eta$ or \etapr resonance;
\item \Ds decays where at least one pion comes from an $\omega$ or $\phi$ resonance;
\item \Ds decays where the three pions come from the \Ds meson with or without an intermediate \aone state:
$\Kz 3\pi$, $\eta 3\pi$,  $\eta^{\prime}3\pi$, $\omega 3\pi$, $\phi 3\pi$, $\tauon\nu_{\tau}$ and nonresonant $3\pi$.
\end{itemize}
The contributions of these \DsTopipipiX decays are estimated from a simultaneous binned maximum-likelihood fit performed on the distribution of four variables: the invariant mass of the $3\pi$ system $m(3\pi)$, the mass of the same-charge pions $m(\pip\pip)$, and the minimum and maximum mass value of the two \pip\pim combinations $\min[m(\pip\pim)]$ and $\max[m(\pip\pim)]$, respectively.
The templates of each \Ds component are described using the simulated inclusive \BToDstDsX samples, while the background from non-\Ds decays is modeled with the simulated inclusive \bbToDstpiX sample.
The free parameters of the fit are the total number of \Ds events, the fractions of the four resonances $\eta$, $\etapr$, $\omega$ and $\phi$, and the relative fractions of the $\eta3\pi$, $\etapr3\pi$ and $\omega3\pi$ decay modes. The parameters related to the other components are constrained or fixed to their values reported in Ref.~\cite{PDG2022}, since their branching fractions are sufficiently well measured.
The results of these fits were reported in detail in Refs.~\cite{LHCb-PAPER-2017-027, LHCb-PAPER-2022-052}.

\subsection{Constraining the \texorpdfstring{\Ds}{Ds} production model}

The different \BToDstDsX decays share similar distributions since the produced excited charm-strange mesons all decay to a \Ds and a soft photon or a \piz and therefore, it is advantageous to constrain their yields in the final fit.
Their relative fractions are obtained from a fit to a specific control sample enriched with these decays and then constrained in the final fit.
The control sample is obtained applying the same signal selection discussed in Section~\ref{sec:selection}, removing the anti-\Ds requirement and the requirement on the maximum \B and \tauon mass values, and requiring the $3\pi$ system to have an invariant mass within 20\mevcc of the known \Ds mass value.
The surviving events are categorized as the exclusive \BdToDstDss, \BdToDstDs, \BdToDstDssz, \BdToDstDspu decays and the inclusive \BduToDstDsX and \BsToDstDsX decays, where the first mode is used as normalization.

\begin{figure}[tb]
\centering
 \includegraphics[width=0.49\textwidth]{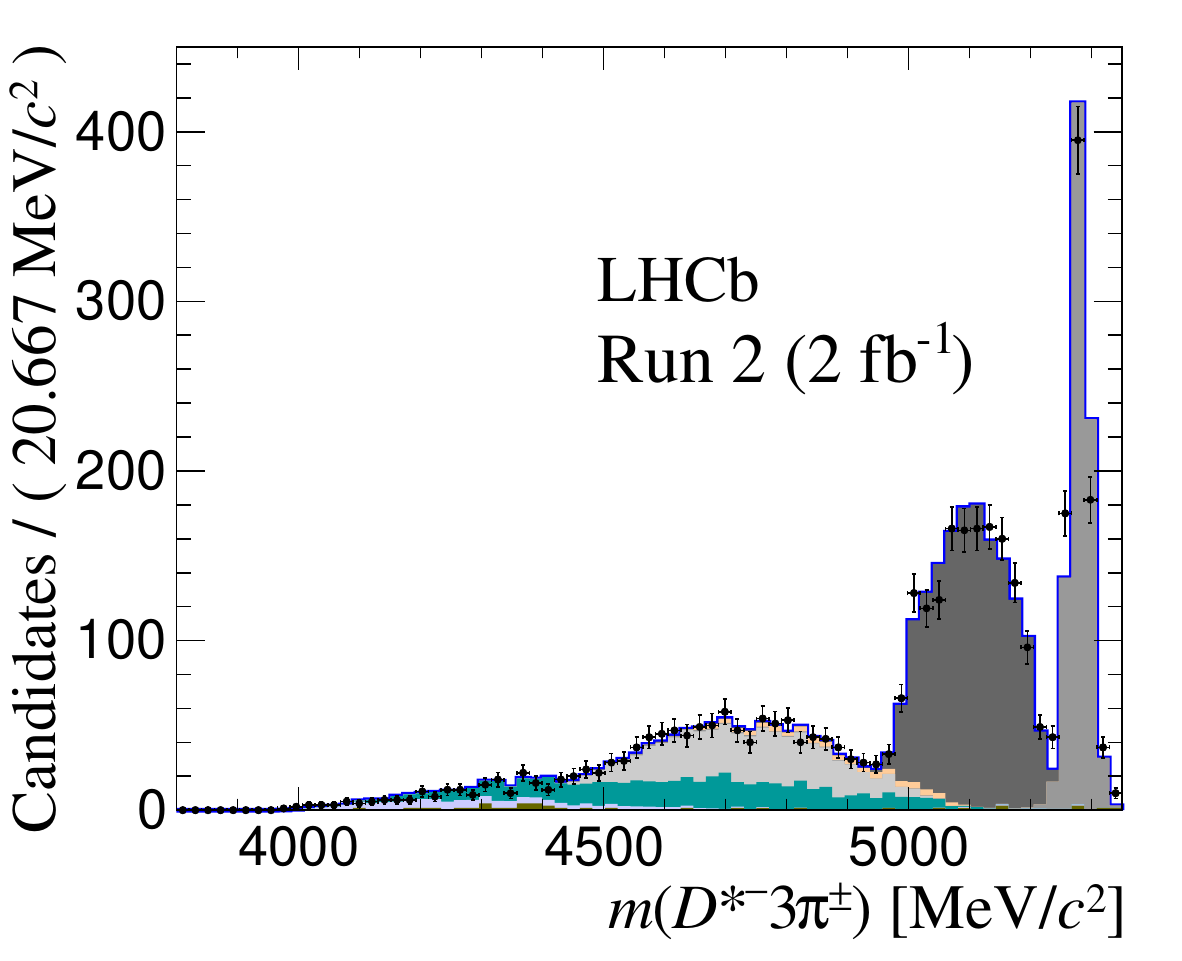}
 \includegraphics[width=0.49\textwidth]{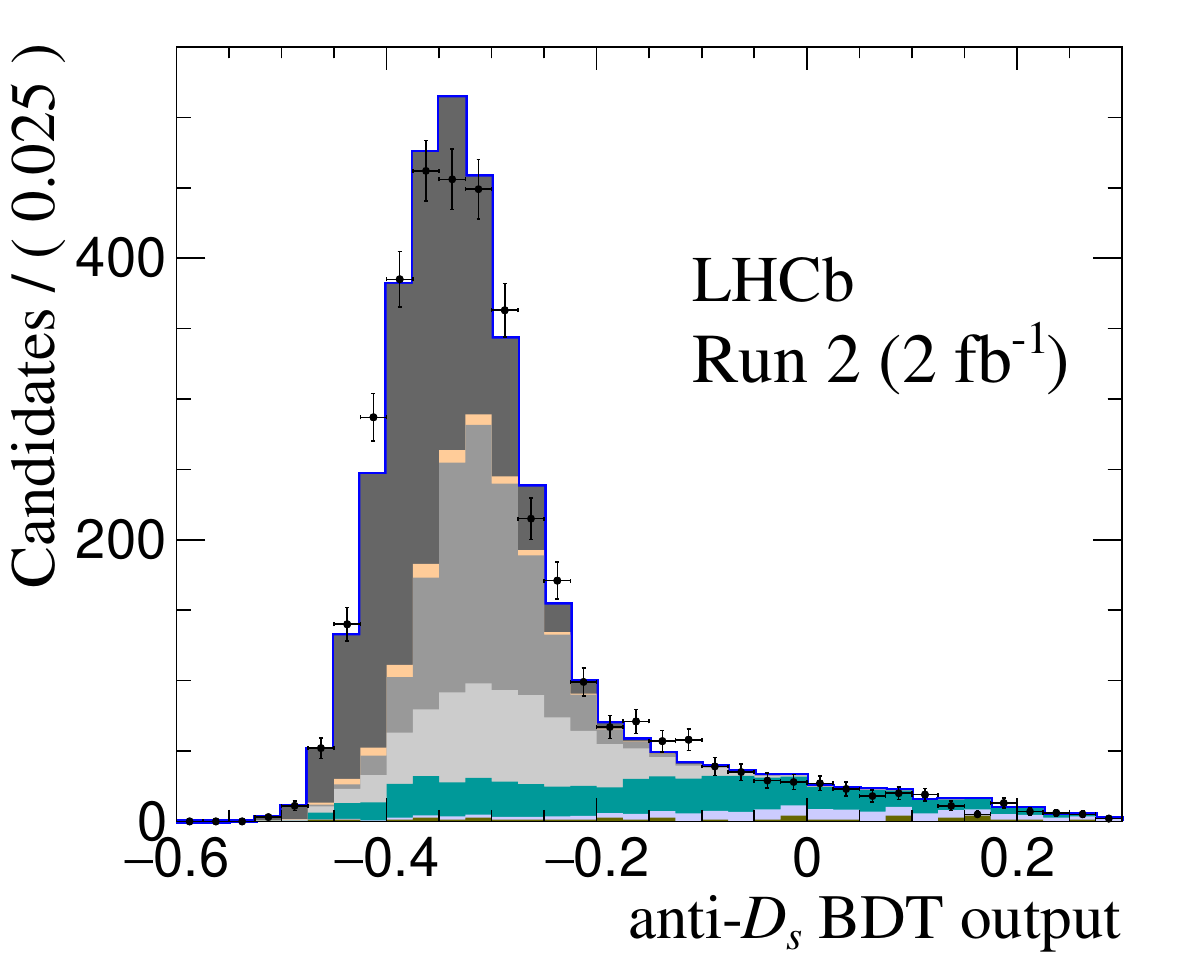}  \\
  \includegraphics[width=0.49\textwidth]{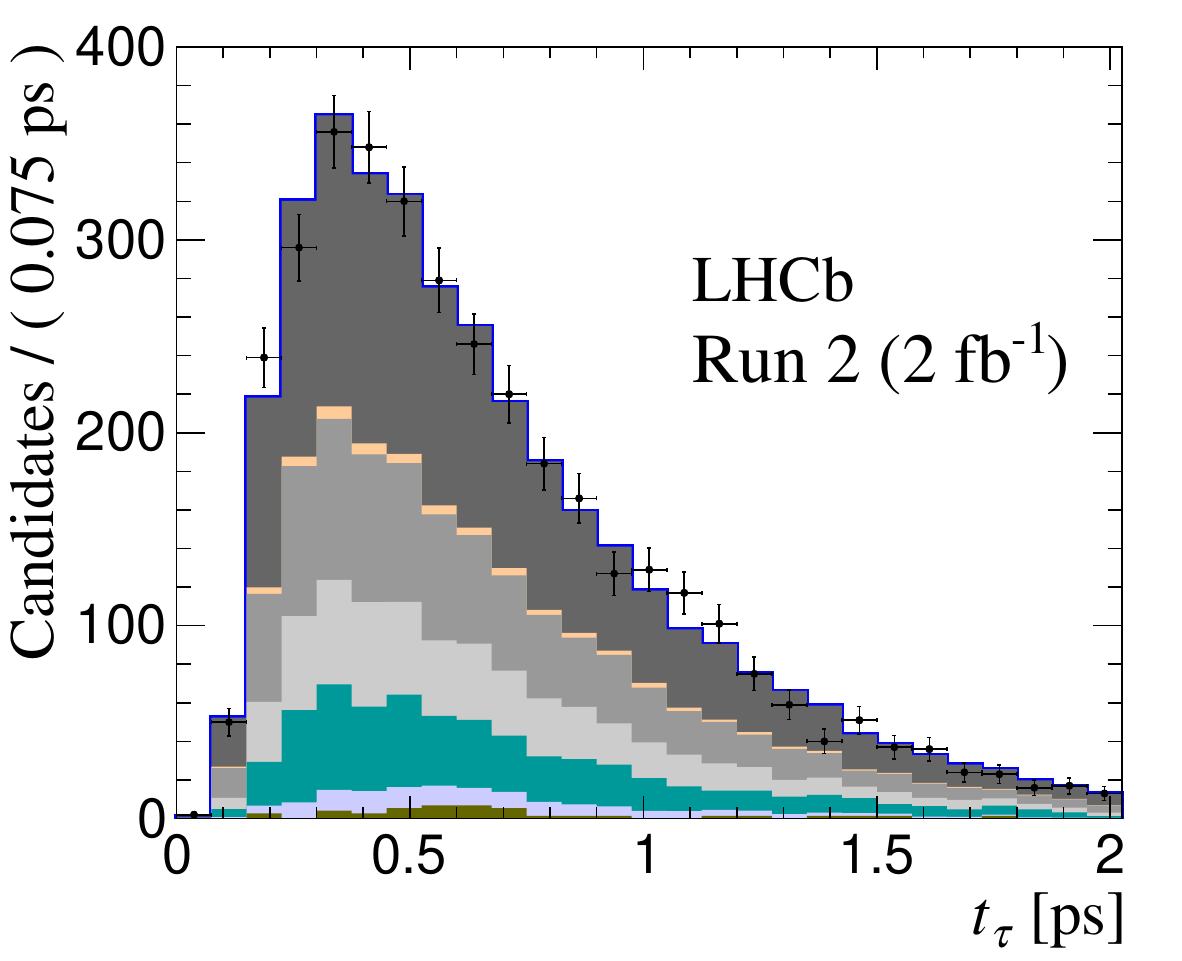}
 \includegraphics[width=0.49\textwidth]{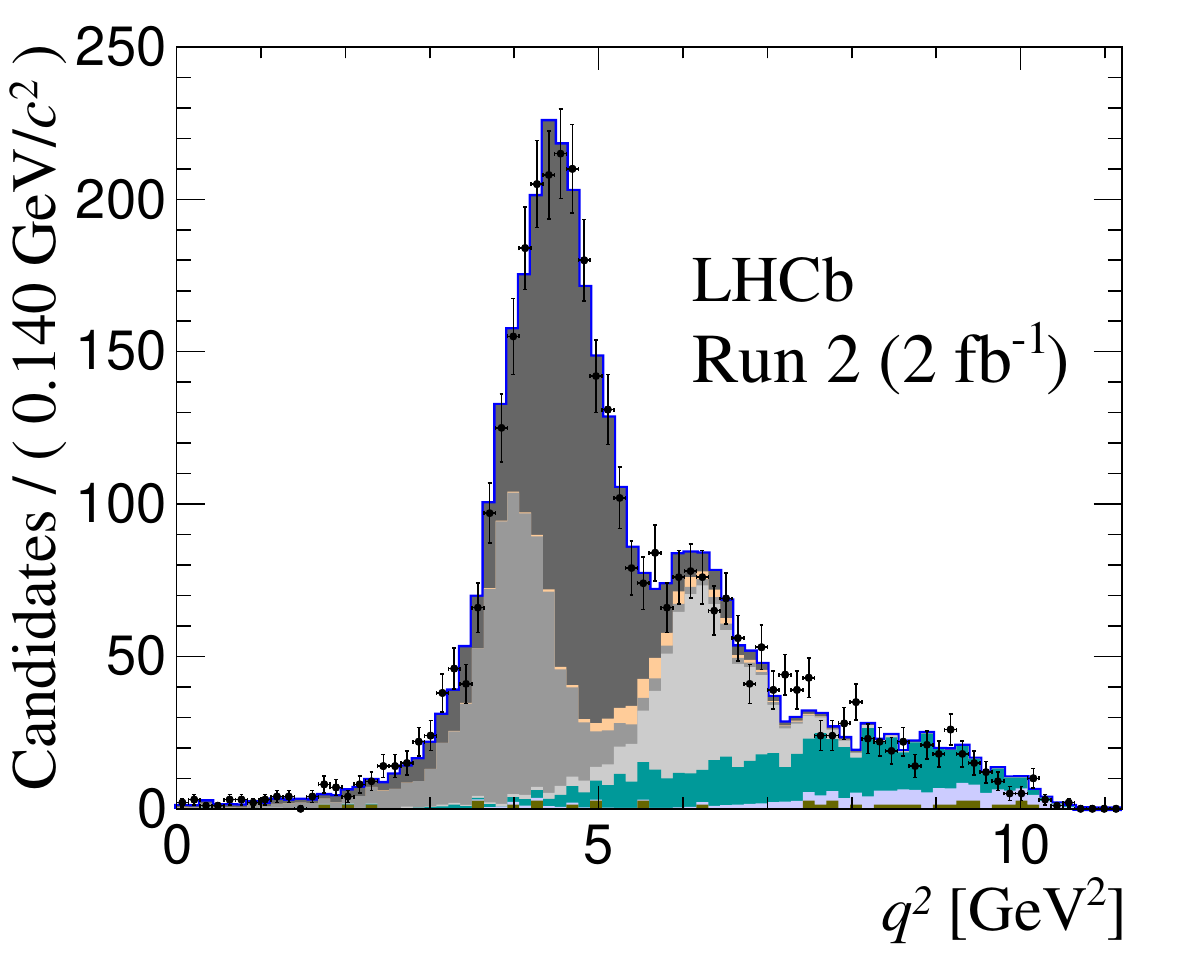}  \\
 \includegraphics[width=0.49\textwidth]{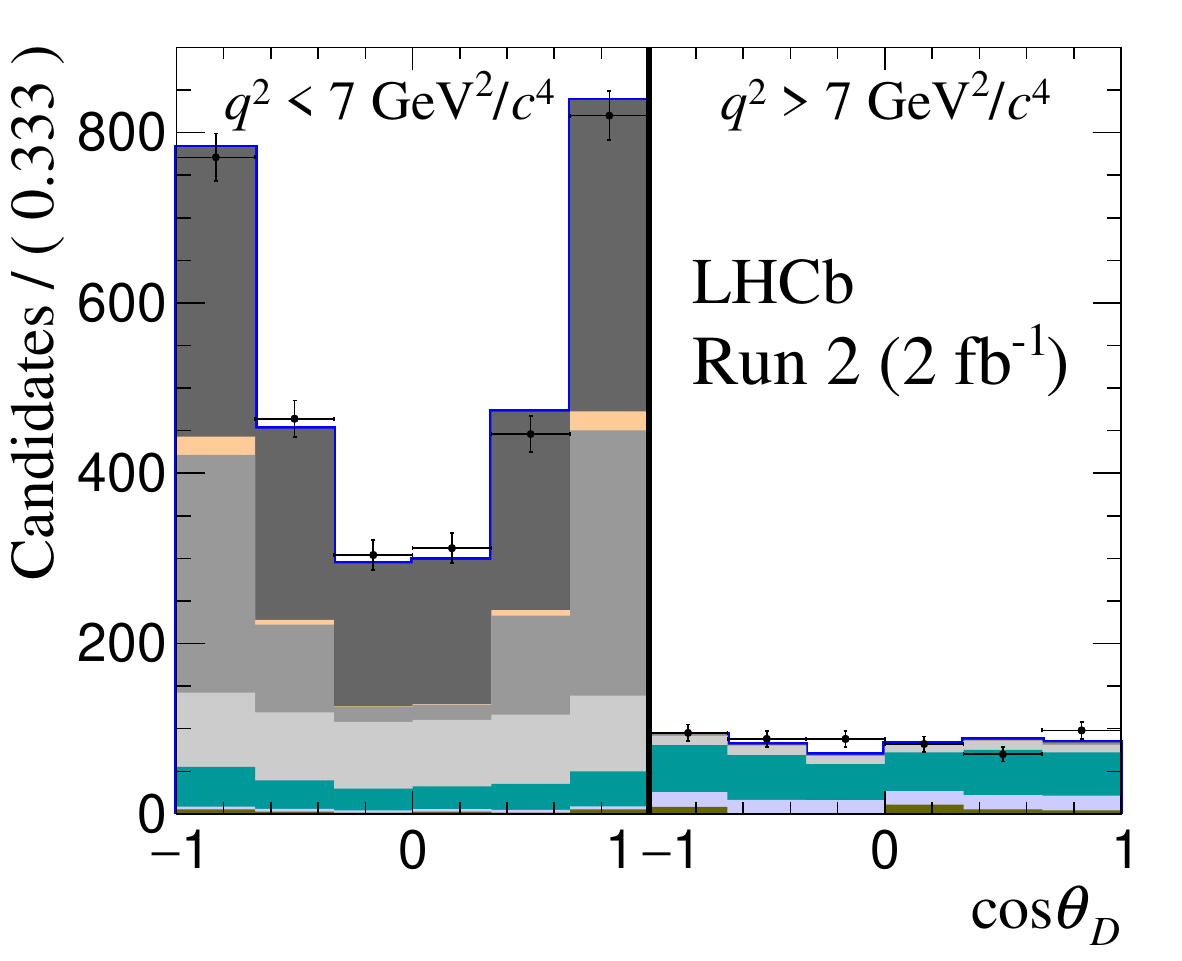} \hskip 3.6cm
  \includegraphics[width=0.25\textwidth]{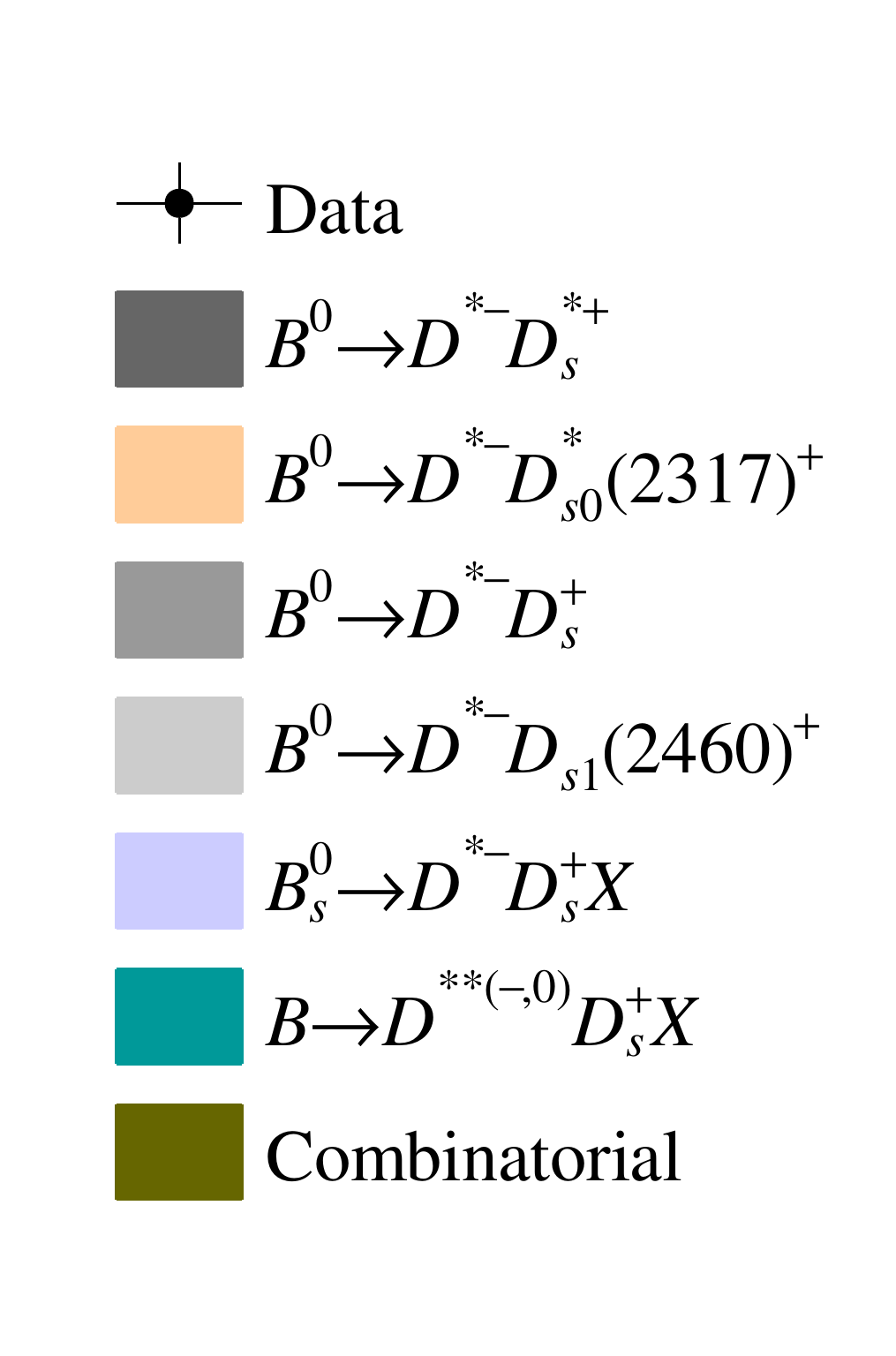} \\
 \caption{Distribution of the key analysis variables obtained in the \BToDstDsX enhanced \runt data sample  with the result of the fit superimposed: (top left) $m(\Dstar 3\pi)$, (top right) anti-\Ds BDT output, (middle left) $t_{\tau}$, (middle right) \qsq and (bottom) \costhetaD.}
    \label{fig:fit_rewDs}
\end{figure}

The \qsq distribution for the exclusive modes peaks exactly at the mass of the \Dsr states, 
while the \qsq for the other modes is higher due to the unreconstructed particles in the final state.
An extended, binned maximum-likelihood fit is performed to the distribution of the $\Dstar3\pi$ mass, where the template of each \Ds component is taken from the simulation and the distribution of the possible combinatorial background is taken from a data sample obtained by combining \Dstar and \tauon candidates with the same sign.
The relative fractions extracted from the fit are reported in Table~\ref{tab:fit_rewDs} while the fit projections on the fitted and other control variables are shown in Figure~\ref{fig:fit_rewDs}.
The central values and the related uncertainties are included as Gaussian constraints on the various \BToDstDsX components in the final fit, accounting for the different efficiencies between control and signal samples.

\begin{table}[tb]
  \centering
  \caption{Decay fractions for the \BToDstDsX decays, normalized for the \BdToDstDss yield, obtained from the \runo and \runt control samples.}
   \begin{tabular}{lcc}
   Parameter & \runo & \runt \\
   \hline
   $f_{\BdToDstDs}$   & 0.56 $\pm$ 0.05 & 0.56 $\pm$ 0.03 \\
   $f_{\BdToDstDssz}$ & 0.01 $\pm$ 0.05 & 0.08 $\pm$ 0.04 \\
   $f_{\BdToDstDspu}$ & 0.37 $\pm$ 0.05 & 0.44 $\pm$ 0.04 \\
   $f_{\BduToDstDsX}$ & 0.36 $\pm$ 0.07 & 0.34 $\pm$ 0.03 \\
   $f_{\BsToDstDsX}$  & 0.07 $\pm$ 0.03 & 0.15 $\pm$ 0.04 \\
   \end{tabular}
 \label{tab:fit_rewDs}
\end{table}

\subsection{Other background sources}
In addition to the double-charm background, the selected sample is also populated by the combinatorial background and the remnant \BToDstpipipiX candidates, surviving the vertex-detachment requirement and the requirement on the \B mass to be lower than 5100\mevcc. 
Nevertheless, these modes are not expected to affect significantly the signal fit because of their mostly unpolarized \costhetaD distribution and the small yield surviving the full signal selection.

\section{Signal fit}\label{sec:fit}

The number of polarized and unpolarized signal events is estimated from an extended maximum-likelihood fit to the four-dimensional distribution of \costhetaD, \qsq, $t_{\tauon}$, anti-\Ds BDT output.
The fit is performed within the SM framework with the free parameters listed below.
\begin{itemize}
\item \Nunpollow and \Nunpolhigh: parameters accounting for the number of unpolarized signal events in the low- and high-\qsq regions, respectively.
\item \fpollow and \fpolhigh: parameters accounting for the fraction of signal
polarized events with respect to the number of unpolarized signal events in the low- and high-\qsq regions, respectively.
\item $N_{\BdToDstDpX}$: parameter accounting for the number of \BdToDstDpX events.
\item $f^{\rm v_1v_2}_{\BdToDstDzX}$: free parameter accounting for the fraction of \BdToDstDzX events where at least 
one pion comes from a different vertex than the \Dz vertex, with respect to $N^{\rm same}_{\BdToDstDzX}$.
\item $N_{\BdToDstDss}$: parameter accounting for the yield of the \BdToDstDss mode.
\end{itemize}

To ensure the stability of the fit some parameters are constrained or fixed:
\begin{itemize}
\item $f_{\tauTopipipiz}$: fraction of \tauTopipipiz signal events with respect to the \mbox{\tauTopipipi} mode. 
This parameter is fixed taking into account the different branching ratios and efficiencies of the two modes.
\item $f_{\BdToDststTauNu}$: fraction of $\BdToDststTauNu$ decays with respect to the $\BdToDstTauNu$ signal. This parameter
is fixed in the fit to the expected value determined from simulation after correcting the branching fractions used in the generation.\footnote{The correction is done by comparing the branching fractions of \BdToDststTauNu~\cite{PDG2022} and the theoretical expectations for each \Dstst state from the $R(\Dstst)$ predictions reported in Ref.~\cite{Bernlochner_2022}.}
\item $f_{\BdToDstDspu}$, $f_{\BdToDstDs}$, $f_{\BduToDstDsX}$, $f_{\BsToDstDsX}$ and $f_{\BdToDstDssz}$:
set of parameters representing the fraction of the relevant decay mode of interest with respect to the \BdToDstDss decay. 
These parameters are constrained to the results of the fit described in Section~\ref{sec:backgrounds} after correcting for efficiency.
\item $N^{\rm same}_{\BdToDstDzX}$: number of \BdToDstDzX events where the three pions in the final state come from 
the same vertex. This parameter is Gaussian constrained to the number of exclusive \DzToKpipipi events
recovered by the isolation tool after correcting for the data-simulation differences. 
\item $N_{\BdToDstpipipiX}$: number of prompt \BdToDstpipipiX events. The central value is determined from the observed ratio between exclusive \BdToDstpipipi and the inclusive \BdToDstpipipiX decays, corrected for data-simulation differences.
\item $N_{\rm B1-B2}$: number of combinatorial backgrounds where the \Dstar and the three pions come from 
different \bquark hadrons. Its value is fixed to the value observed in the wrong-sign sample satisfying the nonisolation and the higher \B-mass requirements.
\item $N_{\rm{fake}\ \textit{\Dz}}$ and $N_{\rm{fake}\ \textit{\Dstar}}$: number of combinatorial background events where a fake \Dz or \Dstar is reconstructed, respectively. Their value is fixed to the values obtained from a fit to $m(K\pi)$ and $m(K\pi\pi)-m(K\pi)$.
\end{itemize}
The fractions \fpol, $f_{\tauTopipipiz}$ and $f_{\BdToDststTauNu}$ are assumed to be the same in the \runo and \runt data samples.

\begin{figure}[!htbp]
    \centering
     \includegraphics[width=0.49\textwidth]{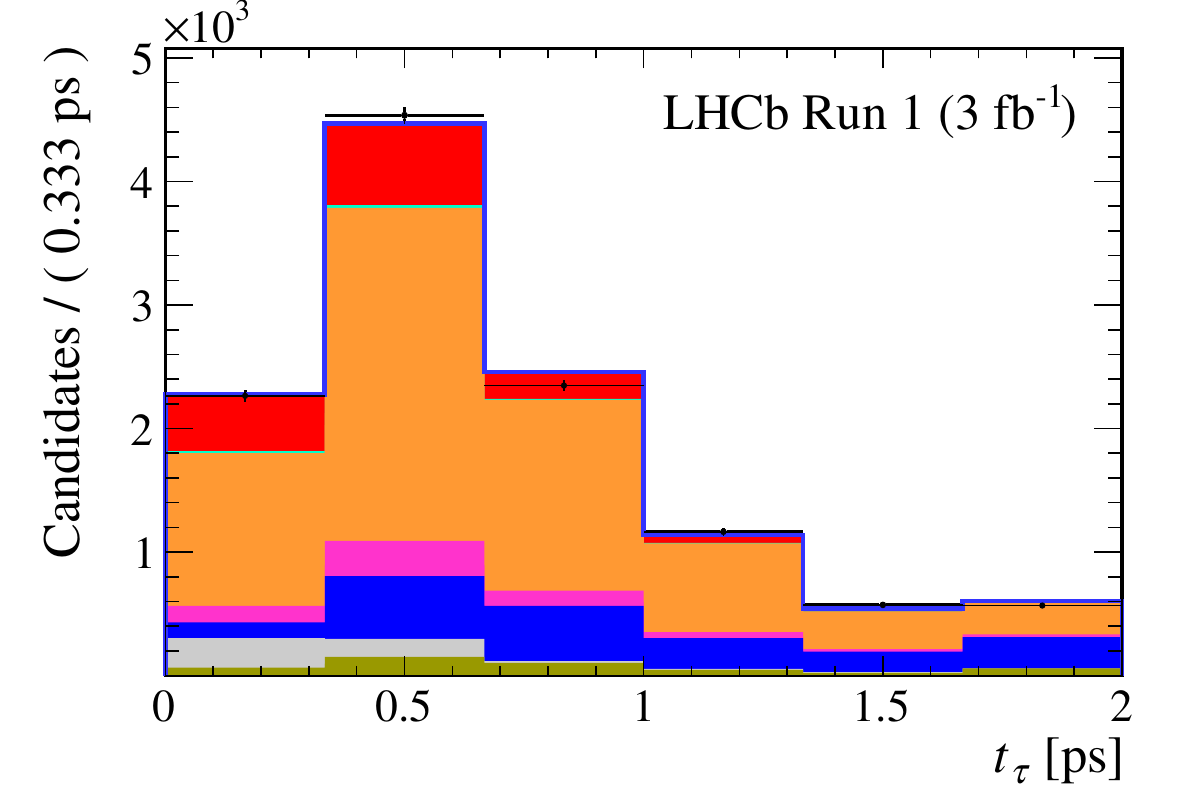}
     \includegraphics[width=0.49\textwidth]{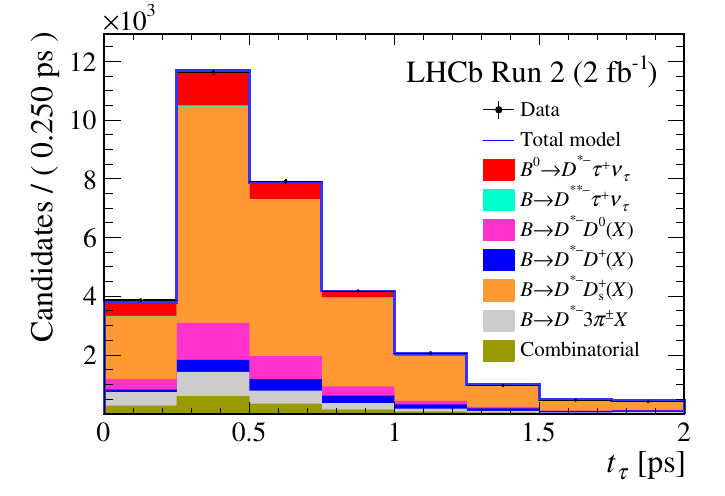} \\
     \includegraphics[width=0.49\textwidth]{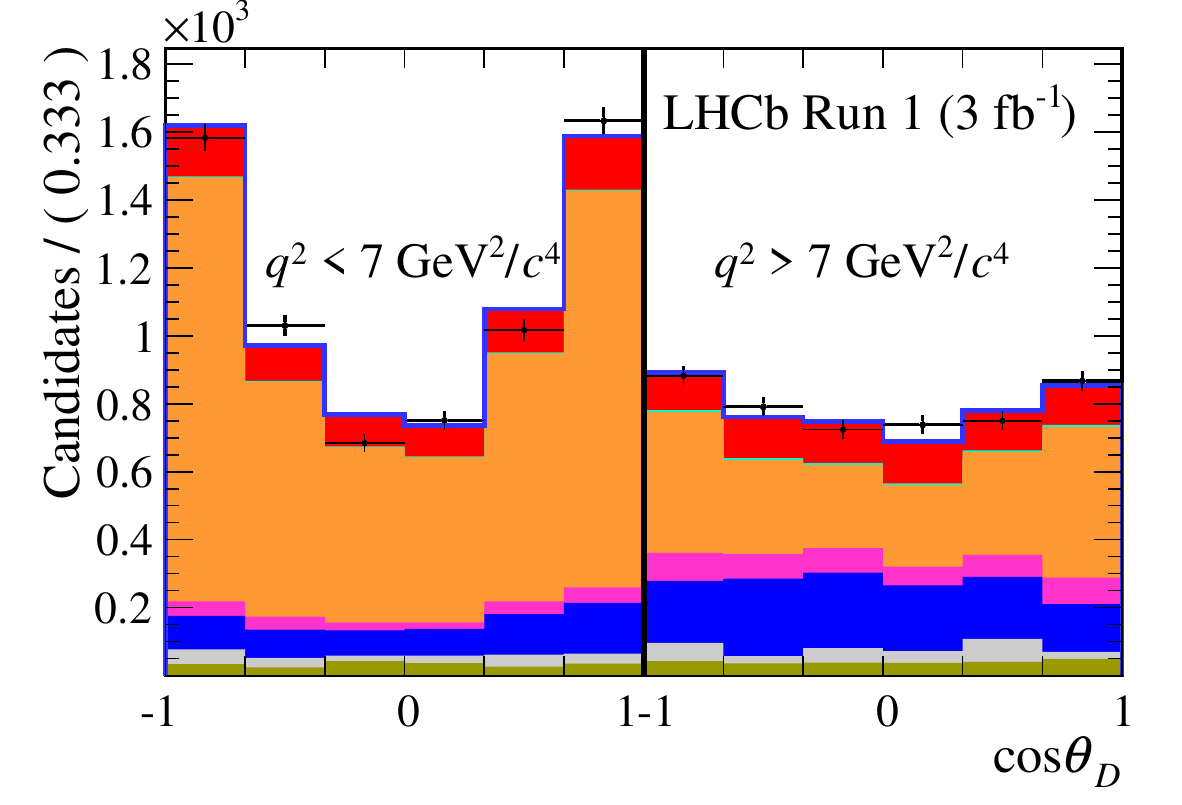}
     \includegraphics[width=0.49\textwidth]{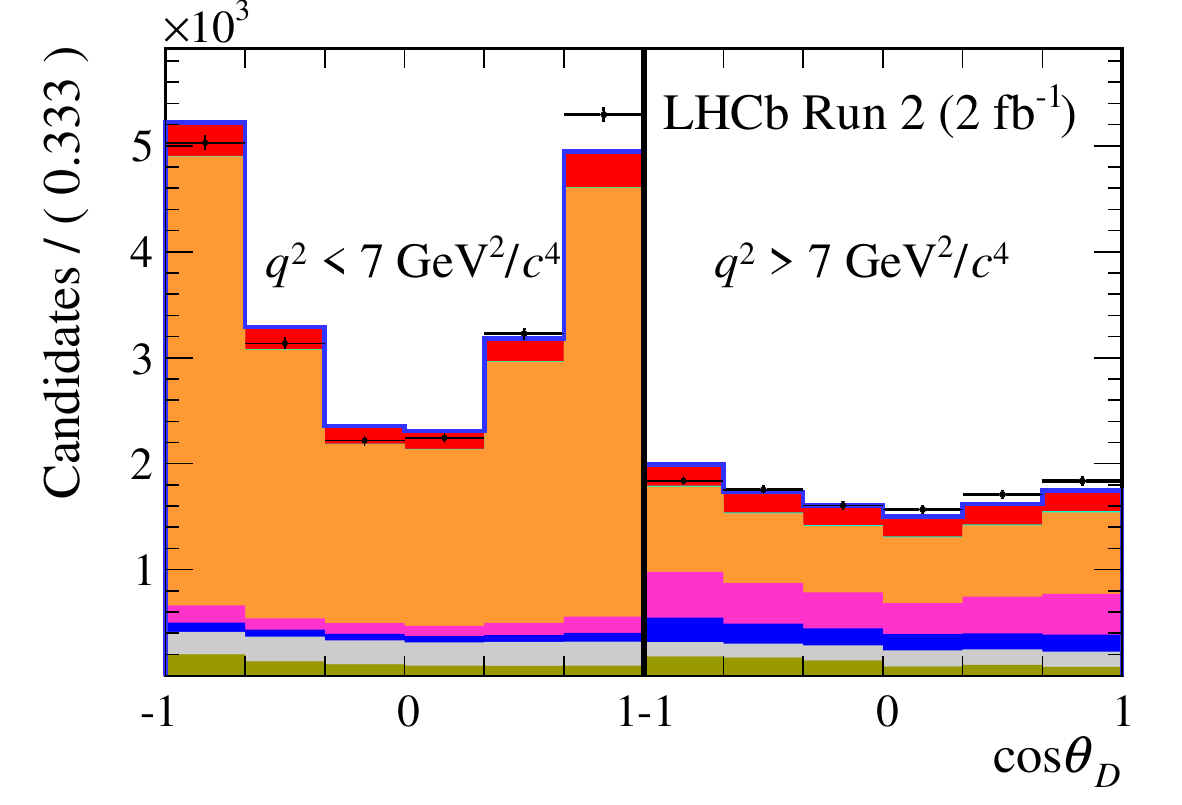} \\
     \includegraphics[width=0.49\textwidth]{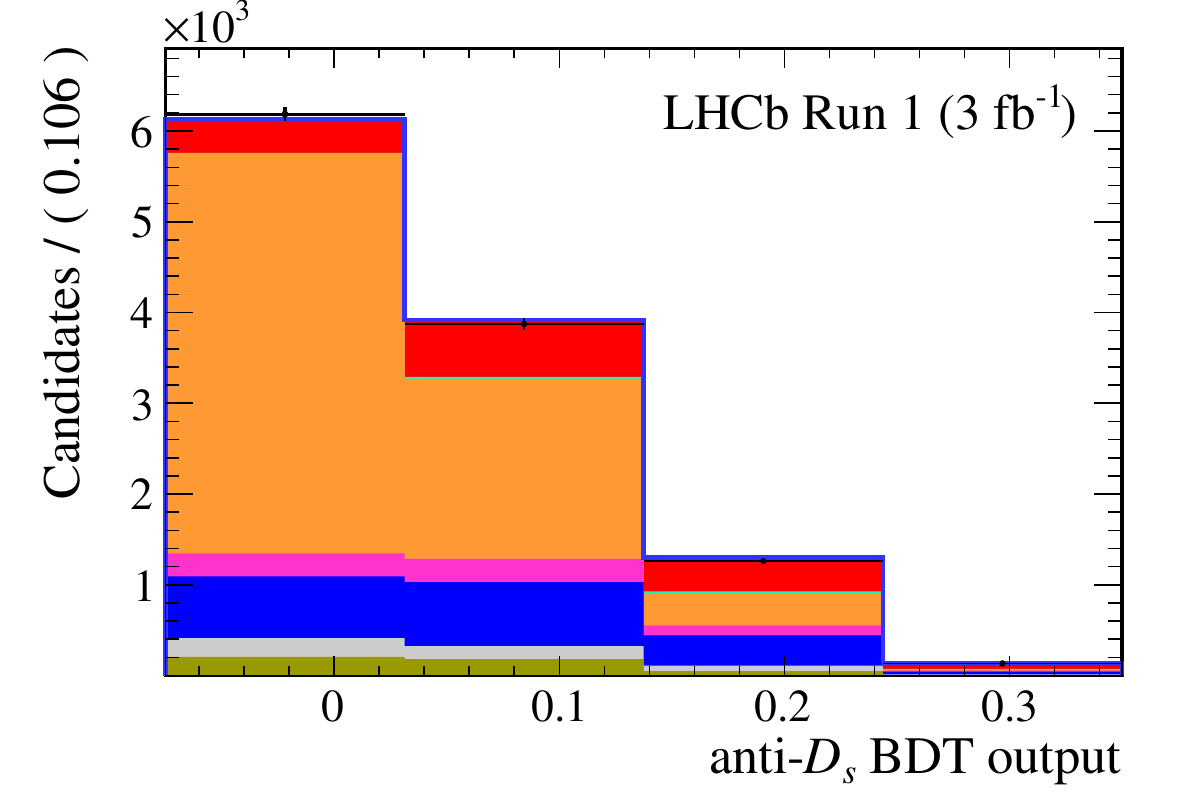}
     \includegraphics[width=0.49\textwidth]{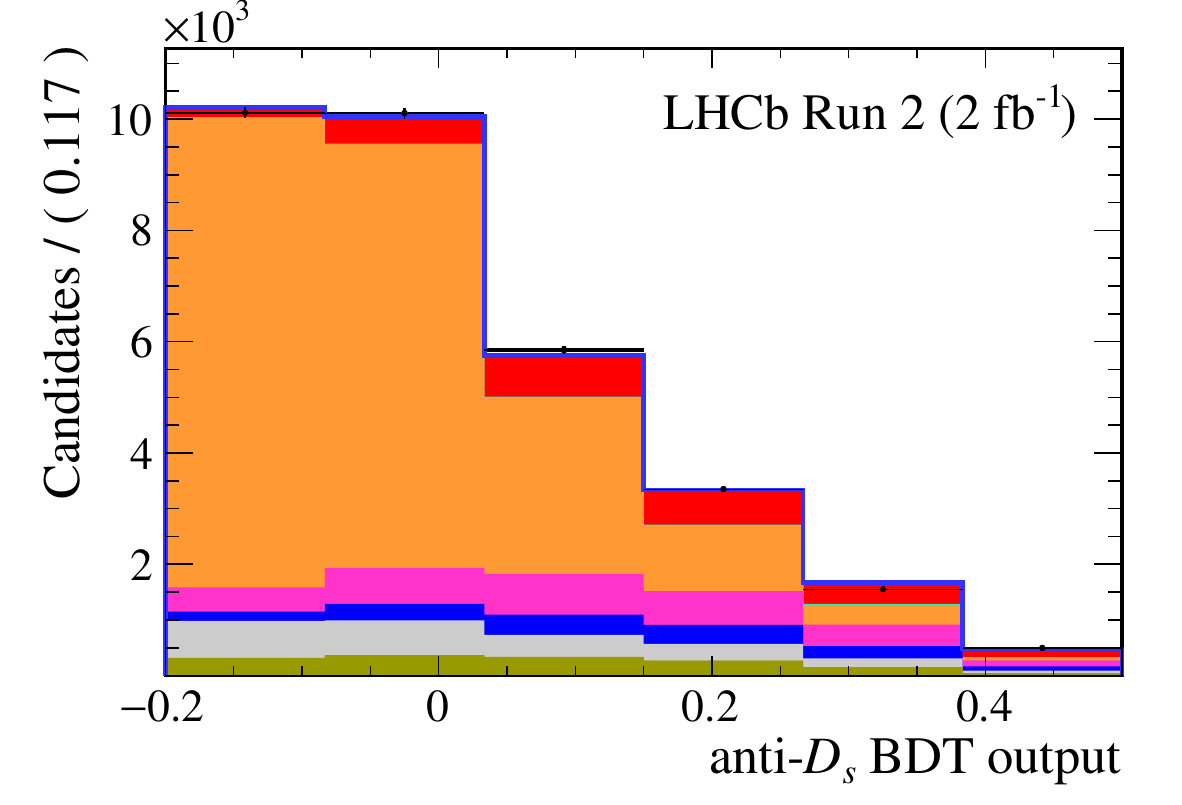} \\
    \caption{Distributions of the fit variables in the (left) \runo and (right) \runt data sets with the fit results superimposed.}
    \label{fig:fit_signal}
\end{figure}

Figure~\ref{fig:fit_signal} shows the distributions of the variables used in the fit, with a $\chi^2$ per degree of freedom of 806/736. The results are summarized in Table~\ref{tab:fit_signal}.
The fractions of polarized signal events are 0.361 $\pm$ 0.074 \stat and 0.013 $\pm$ 0.081 \stat for the low- and high-\qsq regions, respectively. The fit has been repeated on the \runo and \runt samples separately and no significant discrepancies have been observed in all the parameters.

\begin{table}[!htbp]
  \centering
  \caption{Fit results for the \runo and \runt data sets.}
 \begin{tabular}{lcc}
 Parameter & \runo & \runt \\
\hline
\Nunpollow & 360 $\pm$ 55 & 758 $\pm$ 62 \\
\Nunpolhigh & 532 $\pm$ 70 & 827 $\pm$ 109 \\
\fpollow & \multicolumn{2}{c}{0.36 $\pm$ 0.07}\\
\fpolhigh & \multicolumn{2}{c}{0.01 $\pm$ 0.08} \\
$f_{\tauTopipipiz}$  & \multicolumn{2}{c}{0.28} \\
$f_{\BdToDststTauNu}$  & \multicolumn{2}{c}{0.044} \\
$N_{\BdToDstDss}$ & 2087 $\pm$ 77 & 7475 $\pm$ 170 \\
$f_{\BdToDstDspu}$ & 0.38 $\pm$ 0.05 & 0.37 $\pm$ 0.04 \\
$f_{\BdToDstDs}$ & 0.51 $\pm$ 0.03 & 0.60 $\pm$ 0.03 \\
$f_{\BduToDstDsX}$ & 0.83 $\pm$ 0.06 & 0.48 $\pm$ 0.05 \\
$f_{\BsToDstDsX}$ & 0.17 $\pm$ 0.03 & 0.10 $\pm$ 0.02 \\
$f_{\BdToDstDssz}$  & 0.11 $\pm$ 0.02 & 0.02 $\pm$ 0.03 \\
$N^{\rm same}_{\BdToDstDzX}$ & 448 $\pm$ 22 & 1039 $\pm$ 52 \\
$f^{\rm v_1v_2}_{\BdToDstDzX}$ & 0.39 $\pm$ 0.18 & 2.26 $\pm$ 0.28 \\
$N_{\BdToDstDpX}$ & 1747 $\pm$ 118 & 1740 $\pm$ 182 \\
$N_{\BdToDstpipipiX}$ & 408 $\pm$ 21 & 2190 $\pm$ 33 \\
$N_{\rm B1-B2}$ & 197 & 216 \\
$N_{\rm{fake}\ \textit{\Dz}}$ & 110 & 457 \\ 
$N_{\rm{fake}\ \textit{\Dstar}}$ & 133 & 533 \\
\end{tabular}
 \label{tab:fit_signal}
\end{table}

The observed signal yields are corrected for the efficiency mentioned in Section~\ref{sec:signal}.
The \aD and \cD coefficients of the generated \costhetaD distribution are directly related to the efficiency-corrected yields of the polarized and unpolarized signal components in the two \qsq bins. They are determined as
\begin{equation}
\aD = \frac{\Nunpoltrue}{\Ntottrue}\cdot \PDFunpol \vert_{\costhetaD = 0}, \qquad \qquad 
\cD = \frac{3}{2}\frac{\Npoltrue}{\Ntottrue}\Delta,
\end{equation}
where $\PDFunpol\vert_{\costhetaD = 0}$ represents the value of the unpolarized signal PDF evaluated at 0, $\Delta$ is the chosen \costhetaD bin width (0.5), the factor $3/2$ arises from the integral of $\cos^2\theta_D$ distribution,  \Nunpoltrue, \Npoltrue and \Ntottrue 
are the efficiency-corrected signal yields for the unpolarized, polarized and the total PDFs, respectively, and we consider both \runo and \runt samples.
The corresponding \Fld value over the whole \qsq region is determined as the weighted average of the two \Fld values, using the efficiency-corrected signal yield in each \qsq bin as weight.
The extracted \Fld, \aD and \cD values for the binned and total \qsq regions are reported in Table~\ref{tab:fld_results}.

\begin{table}[b]
    \centering 
    \caption{Values of \aD, \cD and \Fld from the fit.}
    \begin{tabular}{ccccccc}
    Parameter & & $\qsq < 7\gevgevcccc$  & & $\qsq > 7\gevgevcccc$ & & Whole \qsq range \\
    \hline
    \aD  & & 0.12 $\pm$ 0.02 & & 0.16 $\pm$ 0.03 & & 0.15 $\pm$ 0.03 \\
    \cD  & & 0.14 $\pm$ 0.05 & & 0.01 $\pm$ 0.02 & & 0.06 $\pm$ 0.06 \\
    \Fld & & 0.52 $\pm$ 0.07 & & 0.34 $\pm$ 0.08 & & 0.41 $\pm$ 0.06 \\
    \end{tabular}
    \label{tab:fld_results}
\end{table}

\section{Systematic uncertainties}\label{sec:systematics}

The main contributions to the systematic uncertainty are due to the signal selection requirements, \qsq bin migration, signal and background modeling, and limited size of the simulated samples.
A breakdown of the systematic uncertainties is reported in Table~\ref{tab:finalSyst} while in the following a brief description is provided for each source.

The simulated signal distributions are weighted according to the form-factor values predicted by the CLN parametrization. Two systematic uncertainties are assigned due to limited knowledge of the form factors: \textit{FF parameters} is evaluated varying the CLN parameters within their uncertainties, while \textit{FF model} is determined repeating the fit with the Boyd-Grinstein-Lebed parametrization~\cite{Boyd_1997,Bernlochner_2017}.

The effects of the \textit{signal selection} and the \qsq \textit{bin migration} on the signal \costhetaD template are also considered. 
A systematic uncertainty is assigned varying these two corrections within their statistical uncertainties.
The presence of any bias in the fit procedure is assessed repeating the fit on a set of pseudoexperiments, obtained with a bootstrap technique from the data distribution (\textit{Fit validation}).
The \Fld value used in the generation of the simulated signal sample (\textit{\Fld in simulation}) can introduce a bias in the reconstruction efficiency and acceptance in the \costhetaD distribution. 
Different \Fld hypotheses are considered in the \Fld range [0.33, 1.0] obtaining each time a new set of signal templates. The fit is repeated and a systematic uncertainty is assigned as the maximum difference with respect to the default \Fld values.

The fraction of \tauTopipipiz ($f_{\tauTopipipiz}$) decays is fixed in the signal fit so a systematic uncertainty is assigned repeating the fit and varying the central value according to its uncertainty. A similar strategy is used for the quantity $f_{\BdToDststTauNu}$, constraining the ratio between the \mbox{\BdToDststTauNu} events and the signal decay (\textit{Fraction of \Dstst feed-down}).

The \Ds decays in simulation are corrected by the weights obtained in the fit to the \DsTopipipiX control sample (\textit{\Ds decay model}).
These weights are varied by taking into account their uncertainties and correlations and the signal fit is repeated with the varied \Ds templates.
Different corrections have been applied to the templates of double-charm backgrounds in order to obtain the data simulation-corrected \costhetaD distributions. For each correction a systematic uncertainty is assigned by varying the corresponding weights within their uncertainty (\textit{Shape of \costhetaD template in $\Dstar\D$ decays}).

Another source of systematic uncertainty is related to the limited size of the simulated sample available for building the fit templates for the various components (\textit{Limited template statistics}). This systematic uncertainty is evaluated repeating the fit using alternative templates obtained after resampling the baseline ones by means of a bootstrap technique. The usage of the ReDecay~\cite{LHCb-DP-2018-004} technique in the generation of large simulated samples helps to keep this effect well under control.

The baseline fit strategy is compared to an alternative method using three-dimensional templates including the \tauon decay time, the anti-\Ds BDT output and \qsq.
The fit results are projected in the \costhetaD variable and a background subtraction is performed. 
Finally the \costhetaD distribution, corrected for reconstruction effects, is fitted using a second-order polynomial for determining  the \aD and \cD parameters.
The results are found to be compatible with the baseline strategy and no systematic uncertainty is assigned.

Finally, an additional systematic uncertainty has been assigned for the determination of the \Fld value in the whole \qsq region. The bin efficiencies used for determining the integrated \Fld value are varied by one standard deviation and the difference with respect to the baseline value is taken as systematic uncertainty (\textit{\Fld integration method}).

\begin{table}[!htbp]
  \centering
  \caption{Summary of the systematic uncertainties related to the \Dstarm polarization measurement.}
 \begin{tabular}{lccc}
Source & Low-\qsq & High-\qsq & Whole \qsq range \\
\hline
Fit validation & 0.003 & 0.002 & 0.003 \\
FF model & 0.007 & 0.003 & 0.005 \\
FF parameters & 0.013 & 0.006 & 0.011 \\
Limited template statistics &  0.026 & 0.017 & 0.019 \\
Fraction of signal \tauTopipipiz decays & 0.001 & 0.001 & 0.001 \\
Fraction of \Dstst feed-down & 0.001 & 0.004 & 0.003 \\ 
Signal selection & 0.005 & 0.004 & 0.005 \\
Bin migration & 0.008 & 0.006 & 0.007 \\
\Fld in simulation & 0.007 & 0.003 & 0.007 \\
\Ds decay model & 0.008 & 0.009 & 0.009 \\
Shape of \costhetaD template in $\Dstarm\Ds$ decays & 0.002 & 0.001 & 0.002 \\
Shape of \costhetaD template in $\Dstarm\Dss$ decays & 0.007 & 0.002 & 0.004 \\
Shape of \costhetaD template in $\Dstarm\Ds X$ decays & 0.007 & 0.006 & 0.007 \\
Shape of \costhetaD template in $\Dstarm\Dp X$ decays & 0.002 & 0.002 & 0.003 \\
Shape of \costhetaD template in $\Dstarm\Dz X$ decays & 0.002 & 0.002 & 0.003 \\
\Fld integration method & - & - & 0.002 \\
\hline
Total & 0.035 & 0.023 & 0.029 \\
\end{tabular}
\label{tab:finalSyst}
\end{table}

\section{Conclusions}\label{sec:results}

A measurement of the \Dstarm longitudinal polarization fraction in \BdToDstTauNu decays was performed by the \lhcb Collaboration using data collected at center-of-mass energies of 7, 8 and 13 \tev and corresponding to an integrated luminosity of 5~\invfb.
The \Fld value was measured in two \qsq bins and in the whole \qsq range, yielding 
\begin{equation*}\label{eq:fld_results}
\begin{split}
    \qsq & < 7\gevgevcccc\ :       \qquad  0.52 \pm 0.07 \stat \pm 0.04 \syst, \\
    \qsq & > 7\gevgevcccc\ :       \qquad  0.34 \pm 0.08 \stat \pm 0.02 \syst, \\
    \qsq &\ \mathrm{whole\ range}: \ \qquad 0.41 \pm 0.06 \stat \pm 0.03 \syst. \\
\end{split}
\end{equation*}
The total correlation between the \Fld values in the two \qsq regions, considering both statistical and systematic effects, was found to be $-0.18$.
These measurements are compatible with SM predictions and with the results obtained by the \belle experiment~\cite{belle_dstarpol}.

%% file: acknowledgements.tex
\section*{Acknowledgments}
%
% These Acknowledgements valid from 3-May-2019
%
\noindent We express our gratitude to our colleagues in the CERN
accelerator departments for the excellent performance of the LHC. We
thank the technical and administrative staff at the LHCb
institutes.
We acknowledge support from CERN and from the national agencies:
CAPES, CNPq, FAPERJ and FINEP (Brazil); 
MOST and NSFC (China); 
CNRS/IN2P3 (France); 
BMBF, DFG and MPG (Germany); 
INFN (Italy); 
NWO (Netherlands); 
MNiSW and NCN (Poland); 
MCID/IFA (Romania); 
%MSHE (Russia); 
MICINN (Spain); 
SNSF and SER (Switzerland); 
NASU (Ukraine); 
STFC (United Kingdom); 
DOE NP and NSF (USA).
We acknowledge the computing resources that are provided by CERN, IN2P3
(France), KIT and DESY (Germany), INFN (Italy), SURF (Netherlands),
PIC (Spain), GridPP (United Kingdom), 
%RRCKI and Yandex LLC (Russia), 
CSCS (Switzerland), IFIN-HH (Romania), CBPF (Brazil),
Polish WLCG  (Poland) and NERSC (USA).
We are indebted to the communities behind the multiple open-source
software packages on which we depend.
Individual groups or members have received support from
ARC and ARDC (Australia);
Key Research Program of Frontier Sciences of CAS, CAS PIFI, CAS CCEPP, 
Fundamental Research Funds for the Central Universities, 
and Sci. \& Tech. Program of Guangzhou (China);
Minciencias (Colombia);
EPLANET, Marie Sk\l{}odowska-Curie Actions, ERC and NextGenerationEU (European Union);
A*MIDEX, ANR, IPhU and Labex P2IO, and R\'{e}gion Auvergne-Rh\^{o}ne-Alpes (France);
%RFBR, RSF and Yandex LLC (Russia);
AvH Foundation (Germany);
ICSC (Italy); 
GVA, XuntaGal, GENCAT, Inditex, InTalent and Prog.~Atracci\'on Talento, CM (Spain);
SRC (Sweden);
the Leverhulme Trust, the Royal Society
 and UKRI (United Kingdom).

%% file: Authorship_LHCb-PAPER-2023-020.tex
% LHCb collaboration author list
% Data extracted on October 21st, 2024 at 3:15pm for paper reference LHCb-PAPER-2023-020
\centerline
{\large\bf LHCb collaboration}
\begin
{flushleft}
\small
R.~Aaij$^{35}$\lhcborcid{0000-0003-0533-1952},
A.S.W.~Abdelmotteleb$^{54}$\lhcborcid{0000-0001-7905-0542},
C.~Abellan~Beteta$^{48}$,
F.~Abudin{\'e}n$^{54}$\lhcborcid{0000-0002-6737-3528},
T.~Ackernley$^{58}$\lhcborcid{0000-0002-5951-3498},
B.~Adeva$^{44}$\lhcborcid{0000-0001-9756-3712},
M.~Adinolfi$^{52}$\lhcborcid{0000-0002-1326-1264},
P.~Adlarson$^{78}$\lhcborcid{0000-0001-6280-3851},
H.~Afsharnia$^{11}$,
C.~Agapopoulou$^{46}$\lhcborcid{0000-0002-2368-0147}
% ...

{\footnotesize \it

% ...
$^{11}$Universit{\'e} Clermont Auvergne, CNRS/IN2P3, LPC, Clermont-Ferrand, France\\
% ...
$^{35}$Nikhef National Institute for Subatomic Physics, Amsterdam, Netherlands\\
% ...
$^{44}$Instituto Galego de F{\'\i}sica de Altas Enerx{\'\i}as (IGFAE), Universidade de Santiago de Compostela, Santiago de Compostela, Spain\\
% ...
$^{46}$European Organization for Nuclear Research (CERN), Geneva, Switzerland\\
% ...
$^{48}$Physik-Institut, Universit{\"a}t Z{\"u}rich, Z{\"u}rich, Switzerland\\
% ...
$^{52}$H.H. Wills Physics Laboratory, University of Bristol, Bristol, United Kingdom\\
% ...
$^{54}$Department of Physics, University of Warwick, Coventry, United Kingdom\\
% ...
$^{57}$School of Physics and Astronomy, University of Glasgow, Glasgow, United Kingdom\\
$^{58}$Oliver Lodge Laboratory, University of Liverpool, Liverpool, United Kingdom\\
% ...
$^{78}$Department of Physics and Astronomy, Uppsala University, Uppsala, Sweden, associated to $^{57}$\\
% ...
}
\end{flushleft}